\documentclass[a4paper,10pt]{article}
\pdfoutput=1
\usepackage{cite}
\usepackage{a4wide}
\usepackage{amsmath}
\usepackage{amsfonts}
\usepackage{amssymb}
\usepackage{graphicx}
\usepackage[left=2.5cm,top=2.2cm,right=2.5cm,bottom=2.6cm]{geometry}

\newcommand{\athdm}[0]{A2HDM }
\newcommand{\athdmws}[0]{A2HDM}
\newcommand{\thdm}[0]{2HDM }
\newcommand{\thdmws}[0]{2HDM}
\newcommand{\cR}{\mathcal{R}}
\newcommand{\eg}{\emph{e.g.}}
\newcommand{\ie}{\emph{i.e.}}
\newcommand{\cO}{\mathcal{O}}

\makeatletter
\def\fmslash{\@ifnextchar[{\fmsl@sh}{\fmsl@sh[0mu]}}
\def\fmsl@sh[#1]#2{%
  \mathchoice
    {\@fmsl@sh\displaystyle{#1}{#2}}%
    {\@fmsl@sh\textstyle{#1}{#2}}%
    {\@fmsl@sh\scriptstyle{#1}{#2}}%
    {\@fmsl@sh\scriptscriptstyle{#1}{#2}}}
\def\@fmsl@sh#1#2#3{\m@th\ooalign{$\hfil#1\mkern#2/\hfil$\crcr$#1#3$}}
\makeatother

\begin{document}
\title{
\begin{flushright}\vbox{\normalsize \mbox{}\vskip -6cm IFIC/13-56\\[-3pt] DO-TH 12/27}%
\end{flushright}
\vskip 30pt
{\bf Electric Dipole Moments in Two-Higgs-Doublet Models}}
\bigskip\bigskip

\author{Martin Jung$^{1}$ and Antonio Pich$^{2}$\\[20pt]
{$^1$\small Institut f\" ur Physik, Technische Universit\" at Dortmund, D-44221 Dortmund, Germany}\\
{$^2$\small IFIC, Universitat de Val\`encia -- CSIC, Apt. Correos 22085, E-46071 Val\`encia, Spain}
}

\date{}
\maketitle
\bigskip \bigskip

\begin{abstract}
\noindent
Electric dipole moments are extremely sensitive probes for additional sources of CP violation in new physics models. Specifically, they have been argued in the past to exclude new CP-violating phases in two-Higgs-doublet models. Since recently models including such phases have been discussed widely, we revisit the available constraints in the presence of mechanisms which are typically invoked to evade flavour-changing neutral currents. To that aim, we start by assessing the necessary calculations on the hadronic, nuclear and atomic/molecular level, deriving expressions with conservative error estimates. Their phenomenological analysis in the context of two-Higgs-doublet models yields strong constraints, in some cases weakened by a cancellation mechanism among contributions from neutral scalars. While the corresponding parameter combinations do not yet have to be unnaturally small, the constraints are likely to preclude large effects in other CP-violating observables. Nevertheless, the generically expected contributions to electric dipole moments in this class of models lie within the projected sensitivity of the next-generation experiments.
\end{abstract}

\newpage

\section{Introduction}
Despite the tremendous success of the Standard Model (SM), the arguments for the necessity of an extension are compelling. Specifically, Sakharov's conditions \cite{Sakharov:1967dj} require the presence of additional CP violation with respect to the SM. 
Assuming CPT invariance, electric dipole moments (EDM) are known to be highly sensitive to new CP-violating phases in new physics (NP) models.
The contributions in the Standard Model are
extremely tiny (\eg{} $d_n^{\rm SM,CKM}\lesssim(10^{-32}-10^{-31})\,e\,{\rm cm}$, see \eg{} \cite{Pospelov:2005pr,Mannel:2012qk} and references therein), with one exception:
the gluonic operator $\mathcal{O}_{G\tilde G}\propto\epsilon_{\mu\nu\rho\sigma}G^{\mu\nu}G^{\rho\sigma}$ gives in principle a contribution many orders of magnitude above the present experimental limits for \eg{} the neutron; this is called the \emph{strong CP problem}. To explain the absence of this contribution, typically symmetries are invoked, involving additional particles. The most famous example is the \emph{Peccei-Quinn mechanism} \cite{Peccei:1977hh}, predicting the presence of axions \cite{Weinberg:1977ma,Wilczek:1977pj}. While these have not yet been found in experimental searches, we implicitly assume in this work when discussing hadronic EDMs that the strong CP problem is solved by this or some similar mechanism. 

The combination of the resulting tiny SM ``background'' and very strong experimental upper limits makes EDMs a well suited laboratory to search for NP, complementary to direct searches at \eg{} the LHC and Tevatron as well as searches involving flavour-changing processes.
The strong suppression in the SM is due to its very specific connection between flavour  and CP violation, \ie{} the Kobayashi-Maskawa mechanism~\cite{Kobayashi}. When new sources of CP violation are included in NP models, usually large contributions are induced, specifically in models which contain flavour-blind phases. Therefore these models include typically an additional mechanism to keep them at bay. This in turn, as realized first by Weinberg \cite{Weinberg:1989dx}, leads in a wide class of models to the situation that the dominant contributions actually stem from two-loop diagrams, when the additional loop allows to avoid strong suppression factors like masses of light quarks or small CKM matrix elements. 

An attractive option for NP is provided by Two-Higgs-Doublet models (\thdmws), 
due to their simplicity and their being the low-energy limit of various theories with a viable UV completion. In the most general version of the model, the fermionic couplings of the neutral scalars are non-diagonal in flavour and, therefore, generate unwanted flavour-changing neutral-current (FCNC) phenomena. Different ways
to suppress FCNCs have been developed, giving rise to a variety of specific implementations of the \thdmws. In the past, mainly \thdmws s without new sources of CP violation have been considered, especially those with a discrete $\mathcal{Z}_2$ symmetry \cite{Glashow:1976nt,Paschos:1976ay}. Recently, however, there has been increased interest in models without this restriction, see \eg{} 
\cite{Cheng:1987rs,Branco:1996bq,Atwood:1996vj,DiazCruz:2009ek,Pich:2009sp,Botella:2009pq,Jung:2010ik,Buras:2010mh,Varzielas:2011jr,Crivellin:2013wna,Shu:2013uua} and also \cite{Branco:2011iw} for a recent review. Potentially huge EDMs used to be the main argument to discard these models. The critical reconsideration of this argument is one of the main motivations for the present work.
We show in this article that while the present experimental limits impose strong bounds on the CP-violating parameter combinations, in models with an appropriate flavour structure they have not yet to be unnaturally small. However, large enhancements in other CP-violating observables are very  strongly restricted by these bounds.
Furthermore, the generic size for EDMs lies well within reach of the next-generation experiments, presently  planned and some already in progress. These will therefore provide critical tests for this class of models in the coming years.

The direct observation of the EDM of a charged particle is very difficult, due to the presence of a hugely  dominating ``monopole'' contribution, \ie{} its charge. Therefore, the most sensitive measurements, at least so far, stem from neutral systems, especially neutrons and atoms/molecules. Relating them to fundamental parameters involves complex calculations at different scales, often implying large uncertainties. Without their careful estimate no reliable constraints on NP parameters can be obtained. We start therefore in the next section by giving model-independent expressions for these observables in terms of Wilson coefficients of the relevant effective operators, taking recent developments into account and estimating the uncertainties of the QCD, nuclear and atomic calculations in a conservative manner. For a subset of systems, this has been done very recently by one of us in \cite{Jung:2013mg}; these results are used when appropriate.
This is followed in Sec.~\ref{sec::exp} by a quick description of the experimental situation, after which we proceed in Sec.~\ref{sec::edmsin2hdms} to discuss the situation of EDMs in \thdmws s with new sources of CP violation. We start by describing the various sources, 
pointing out their different importance. To be specific, we then calculate the resulting constraints in the Aligned Two-Higgs-Doublet Model (\athdmws), which has been introduced in \cite{Pich:2009sp,Jung:2010ik} and whose phenomenology has been further discussed in \cite{Jung:2010ab,Jung:2012vu,Celis:2012dk,Celis:2013rcs,Duarte:2013zfa}. However, the structure of the model is such that the results hold rather generally. In this context, we point out a general cancellation mechanism for neutral scalar contributions, which questions the way they are commonly treated in the literature.
We analyze the phenomenological constraints coming from the presently available experimental bounds in Sec.~\ref{sec::phenomenology}, before giving our conclusions in Sec.~\ref{sec::conclusions}.

\subsection{Comparison to existing work}
There is a huge amount of  literature on EDMs, and there is no hope of reviewing it here; instead, we refer the reader to \cite{Ginges:2003qt,Pospelov:2005pr,Raidal:2008jk,Fukuyama:2012np,Engel:2013lsa} for recent reviews. 
Generally, most of the analyses in the literature are performed within the framework of supersymmetric models (SUSY) (for recent examples, see \cite{Ellis:2008zy,Ellis:2011hp,Hisano:2008hn,Hisano:2006mj,Altmannshofer:2009ne,Li:2010ax,Mercolli:2009ns,Ilakovac:2013wfa,Dhuria:2013ida} and also the phenomenological analysis in \cite{Raidal:2008jk}). While in principle the \thdm  contributions are present in these models as well, they are usually subdominant, which is why they do not receive much attention. Especially the charged Higgs exchange is usually negligible in these models, as it does not exhibit the strong $\tan\beta$-enhancement of other terms, which is why some of the corresponding contributions discussed below are not incorporated at all in these analyses.

Recent studies more closely related to our work include \cite{Trott:2010iz,Buras:2010zm,Batell:2010qw}. In the first of these, the authors discuss one contribution discussed below, namely the charged Higgs contribution to the neutron EDM. The results are similar to ours\footnote{The interested reader can compare them using the relations $\eta_u=\varsigma_u$ and $\eta_d=-\varsigma_d^*$.}, apart from a different treatment of the hadronic matrix element, which yields weaker constraints in our case. The second article discusses EDM contributions in the context of Minimal Flavour Violation (MFV), including complex phases in that framework. The authors perform the analysis in the decoupling limit and assume a small breaking of the $\mathcal{Z}_2$ symmetry, as was assumed already for the \thdm analysis in \cite{D'Ambrosio:2002ex}. Their results are therefore relevant for a subset of our parameter space. They conclude, as we will below, that one-loop contributions are generally not exceeding the experimental limits. In addition, they consider a subset of the two-loop contributions we discuss below, corresponding to the more restrictive assumptions they make.
Finally, in \cite{Batell:2010qw} the authors discuss a subset of MFV operators which might generate a new phase in $B_s$ mixing; the corresponding  operators are not relevant in our context.

\section{Model-independent expressions for EDMs\label{sec::MIexpressions}}
From the point of view of particle physics, the proper starting point for a model-independent analysis is the following effective Lagrangian at the hadronic scale (here up to dimension six, see \eg{} \cite{Pospelov:2005pr}):
\begin{eqnarray}\label{eq::Leff}
\mathcal{L} = -\!\!\sum_{f}\left[\frac{d_f^\gamma}{2}\mathcal{O}_f^\gamma+\frac{d_f^C}{2}\mathcal{O}_f^C\right]+C_W \mathcal{O}_W+\sum_{f,f'}C_{ff'}\mathcal{O}_{ff'}\,,
\end{eqnarray}
with the operator basis
\begin{align}
\mathcal{O}_f^\gamma &= ie\bar{\psi}_f F^{\mu\nu}\sigma_{\mu\nu}\gamma_5\psi_f\,,&\quad \mathcal{O}_f^C &= ig_s\bar{\psi}_f G^{\mu\nu}\sigma_{\mu\nu}\gamma_5\psi_f\,,\nonumber\\
\label{eq::Odef}
\mathcal{O}_W&=+\frac{1}{6}f^{abc}G_{\mu\nu}^a\epsilon^{\nu\beta\rho\sigma}G_{\rho\sigma}^b G_\beta^{\phantom{\beta}\mu,c}\,,&\quad\mathcal{O}_{ff'}&=(\bar{\psi}_f\psi_f)(\bar{\psi}_{f'}i\gamma_5\psi_{f'})\,.
\end{align}
The operators in Eqs.~\eqref{eq::Odef} are the (colour--)EDM operators $\mathcal{O}^{\gamma,C}_f$ for light fermions ($f=e,d,u,s$), the Weinberg operator $\mathcal{O}_W$ and T- and P-violating four-fermion operators $\mathcal{O}_{ff'}$ without derivatives (see, \eg, \cite{Khriplovich:1997ga}).
The factors of $1/2$ for the (C)EDM operators are included to identify the coefficients $d_f^{\gamma,C}$ with the classical electric/gluonic dipole moment in the corresponding limit. The analysis of their influence on experimental observables is divided into two steps: first, the observables have to be expressed in terms of the coefficients of this effective lagrangian. This step can be done independently from the NP model considered and is performed in this section. 
The necessary calculations are on the QCD, nuclear, and/or atomic/molecular level. They typically involve relatively large uncertainties.  
Their careful assessment is essential to obtain reliable bounds, which is why we will pay close attention to this. In the second step, performed exemplarily for the \athdm later in this paper, the coefficients have to be calculated in terms of parameters of the NP model considered, allowing to obtain the constraints on the latter.

In calculations on the QCD level, 
the corresponding matrix elements are often  known only up to a factor of a few, sometimes without a definite sign. There are different methods to calculate/estimate them; while \emph{Naive dimensional analysis} (NDA) \cite{Manohar:1983md} is still used occasionally, mostly due to its simplicity, its estimates are known to be uncertain \eg{} by arbitrary powers of $4\pi$ (see \eg{} \cite{Bigi:1991rh}), which is why we do not consider these estimates here. Instead we are going to use QCD sum rule estimates, where such factors are absent and which are supposed to be uncertain ``only'' by the aforementioned factor of a few (depending on the operator). The main reason for this limited precision is that for sum rules with baryons the suppression of excited states does not work as well as for mesons. For a review on these issues, see \cite{Pospelov:2005pr}. While ultimately progress may come from Lattice QCD, there are severe difficulties obtaining reliable results at the moment, such that we are not aware of available results competitive to the ones used here. Note also available calculations 
in the framework of Baryon Chiral Perturbation Theory \cite{Bernard:2007zu} (see \eg{} \cite{deVries:2010ah,Engel:2013lsa} for recent analyses and references therein). There, the scaling of the various matrix elements can be analyzed from the chiral properties of the operators, leading to a systematic classification scheme. However, it is typically accompanied by NDA estimates, as unknown low-energy constants prevent quantitative estimates. We therefore do not use their results here quantitatively.

The involved calculations on the nuclear and atomic/molecular level are in very different shape, uncertainties ranging from a few to several hundred percent;  
these are commented upon in the appropriate subsections. 
These relatively large uncertainties are in a sense of minor importance for the experimental searches, since NP contributions can easily be larger than the SM ones by several orders of magnitude. However, they are essential in obtaining bounds for NP parameters from the experimental limits.
Readers not interested in their detailed discussion find the final expressions for the corresponding EDMs in Eqs.~\eqref{eq::nEDM},\eqref{eq::dnCW} and~\eqref{eq::HgEDM}. 
For paramagnetic systems, we use the results from \cite{Jung:2013mg}, but perform an update to include the very recent measurement with thorium monoxide (ThO)~\cite{Baron:2013eja}; its results are summarized in Table~\ref{tab::delimitnew}.

\subsection{The neutron EDM}
The neutron EDM can be related to the coefficients in Eq.~\eqref{eq::Leff} by QCD calculations alone.
Here we collect the necessary formulae, for details see again \eg{} the review \cite{Pospelov:2005pr}.  
This EDM is dominated by contributions from the (C)EDMs of its constituents and the Weinberg operator, while four-quark operators play a minor role.

The QCD sum rule calculation for the contribution from the quark (C)EDMs yields \cite{Pospelov:2000bw,Hisano:2012sc}
\begin{eqnarray}\label{eq::nEDM}
d_n\!\left(d_q^{\gamma},d_q^C\right)/e &=& \left(1.0^{+0.5}_{-0.7}\right)\;\left[1.4\left(d_d^\gamma(\mu_h)-0.25\, d_u^\gamma(\mu_h)\right)\right.\nonumber\\
&&\hskip 1.45cm +\,
1.1\left.\left(d_d^C(\mu_h)+ 0.5\, d_u^C(\mu_h)\right)\right]\;\frac{\langle\bar{q}q\rangle(\mu_h)}{(225~{\rm MeV})^3}\,,
\end{eqnarray}
where $\mu_h\sim 1~{\rm GeV}$ denotes a hadronic scale.
In the following we suppress the scale dependence in the notation for brevity and evaluate at $\mu_h=1~{\rm GeV}$ unless stated explicitly.
The uncertainty given here for these matrix elements is similar to the estimate given in \cite{Pospelov:2000bw}. However, given the results in \cite{Hisano:2012sc}, we extended the range to include lower values.\footnote{Note that the analytical differences have minor numerical impact.} This incorporates larger values for the normalization factor $\lambda_n$, determined by the matrix element of the nucleon and its interpolating current, see \cite{Aoki:2008ku,Braun:2008ur,Gruber:2010bj}.\footnote{Note, however, that the rather large central value in \cite{Aoki:2008ku} does lead to a too small value for the nucleon sigma term $\sigma_{\pi N}$ when using the sum rule in \cite{Jin:1993nn} \cite{Ritzprivcomm}.}  We note that alternative treatments are compatible within the estimated level of precision, however indicating in some cases higher sensitivity, see \eg{} \cite{Fuyuto:2012yf}.
Note furthermore that the quark condensate $\langle \bar qq\rangle$ in this formula combines with the light quark masses in the Wilson coefficients as \cite{GellMann:1968rz}
\begin{equation}
(m_u+m_d)\langle\bar{q}q\rangle=-f_\pi^2 m_\pi^2+\mathcal{O}(m_{u,d})\,,
\label{eq::qc}
\end{equation}
which reduces the corresponding uncertainty.	

For the Weinberg operator, the 
contribution reads \cite{Demir:2002gg}\footnote{Here and in Eq.~\eqref{eq::Cbd} the authors state a $~100\%$ uncertainty for the result, which we incorporate as allowing for twice and half the computed value.}
\begin{equation}\label{eq::dnCW}
|d_n(C_W)/e| = \left(1.0^{+1.0}_{-0.5}\right)\,20~\mbox{MeV}\,C_W\,,
\end{equation}
with the sign left undetermined. This expression is based on several estimates that all lead to similar results, but is not a direct calculation.

Finally, for the sake of completeness, for an exemplary four-quark contribution to the neutron EDM the sum rule estimate results in \cite{Demir:2003js}
\begin{equation}\label{eq::Cbd}
|d_n(C_{bd})/e|=2.6\,\left(1.0^{+1.0}_{-0.5}\right)\times10^{-3}~{\rm GeV}^2\left(\frac{C_{bd}(\mu_b)}{m_b(\mu_b)}+0.75\,\frac{C_{db}(\mu_b)}{m_b(\mu_b)}\right)\,,
\end{equation}
again with an unspecified sign. Note that here four-fermion operators involving the beauty quark, defined analogously to their equivalents with light fermions, contribute below the $b$-quark mass scale effectively via an effective two-gluon coupling of the down quark, which is also why the coupling is to be evaluated at $\mu_b\sim m_b$.
The contribution from up-type quarks is ignored, as enhanced couplings in that sector (corresponding \eg{} to $\tan\beta\ll1$ in a Type II model or to $|\varsigma_u|\gg 1$ in the \athdmws) are usually excluded.

\subsection{EDMs of atoms\label{sec::atomEDMs}}
For atoms, Schiff's theorem \cite{Schiff:1963zz} implies a vanishing EDM in the non-relativistic limit for systems of particles whose charge distribution is identical to their EDM distribution. The limits from the non-observation of these EDMs are then related to violations of the conditions for this theorem, and separated into two classes, depending on which of the approximations is more strongly violated. For reviews on atomic calculations, see \eg{} Refs.~\cite{Khriplovich:1997ga,Ginges:2003qt}.

In paramagnetic atoms, \ie{} atoms with non-vanishing total angular momentum, relativistic effects are important, which are largely enhanced for atoms with a large proton number \cite{Sandars:1965xx,Sandars:1966xx,Flambaum:1976vg}, scaling at least like $d\sim Z^3$. This implies a sensitivity mainly to the electron EDM, but also electron-nucleon interactions are enhanced, described by
\begin{equation}\label{eq::HeN}
\mathcal{H}_{eN} = \frac{G_F}{\sqrt{2}}\sum_{N=n,p}\left(\tilde{C}_S^N (\bar{N}N)(\bar{e}i\gamma_5e)+\tilde{C}_P^N (\bar{N}i\gamma_5N)(\bar{e}e)+
\tilde{C}_T^N(\bar{N}i\gamma_5\sigma^{\mu\nu}N)(\bar{e}\sigma_{\mu\nu}e)\right)\,.
\end{equation}
The coefficients of both classes of contributions are estimated in atomic multi-body calculations.
In some publications, these operators are classified instead according to their isospin, 
\begin{equation}\label{eq::CXisospin}
\mathcal{H}_{eN}^X=\sum_N \left[\bar{N}\Gamma_X^1\left(C_X^{(0)}+ C_X^{(1)}\tau_3\right)N\right]\left(\bar{e}\Gamma_X^2e\right)\,,
\end{equation}
where $X=S,P,T$ and the Dirac structures $\Gamma_X^{1,2}$ can be read off from Eq.~\eqref{eq::HeN}.

In diamagnetic atoms, where the total angular momentum vanishes, the finite size of the nucleus is the main source for the violation of Schiff's theorem. The dominant contribution to the corresponding EDM stems from its nuclear \emph{Schiff moment}, which can be expressed in terms of the nucleon EDMs and pion-nucleon couplings, which are in turn  related to the basic terms in Eq.~\eqref{eq::Leff}. Although the quark CEDMs typically give the dominant contribution, the above electron-nucleon interaction is relevant as well.

For  an atom with proton number $Z_p$, neutron number $Z_n$ and consequently nucleon number $A=Z_n+Z_p$, 
the parameter combinations effectively contributing to the EDMs read
(see again \eg{} \cite{Khriplovich:1997ga,Ginges:2003qt})
\begin{equation}
A\tilde C_S\equiv Z_p\tilde C_S^p+Z_n\tilde C_S^n\quad\mbox{and}\quad  \langle\sigma\rangle_{at}\tilde C_{P,T}^{at}\equiv \langle\sigma_n\rangle_{at}\tilde C_{P,T}^n+\langle\sigma_p\rangle_{at}\tilde C_{P,T}^p\,,
\end{equation}
where $\langle\boldsymbol{\sigma}_N\rangle_{at}$ denotes the sum over the spin of the indicated nucleon species in the corresponding nuclear state,
and we used $\langle\boldsymbol{\sigma}\rangle_{at}=\langle\boldsymbol{\sigma}_n\rangle_{at}+\langle\boldsymbol{\sigma}_p\rangle_{at}$ and $\langle\boldsymbol\sigma_i\rangle_{at}=\langle\sigma_i\rangle_{at}\mathbf{I}/I$, where $\mathbf{I}$ denotes the total nuclear spin. The spin sums stem from the quantum-mechanical expressions derived from the pseudoscalar operator $\bar N\gamma_5 N$.
In this equation, the contribution from the first term in Eq.~\eqref{eq::HeN} is seen to be additionally enhanced, because the contributions from neutrons and protons enter spin-independently. This renders this term dominant for paramagnetic systems, as for the other two coefficients closed shells in the nucleus barely contribute. In diamagnetic atoms, it does not contribute at leading order, however, which is why the relative influence of the other two terms is relatively enhanced. 
In fact, if present, among the electron-nucleon interactions the third term is typically dominant in this case.

In general, the definitions for $\tilde C_X^{(at)}$ imply a dependence of these coefficients on the system considered. However, because of $(Z_n+Z_p)/A=1$ and $\tilde C_S^n\approx \tilde C_S^p$, this is usually neglected in the case of $\tilde C_S$. More importantly, the ratios $Z_N/A$ are approximately universal for the systems considered here, leading to a universal $\tilde C_S$ even for $\tilde C_S^n\neq \tilde C_S^p$ \cite{Jung:2013mg}. However, for the spin-dependent terms the relative weights are not atom-independent, such that $\tilde C_{P,T}^{at}$ depend on the atom if $\tilde C_{P,T}^n\neq \tilde C_{P,T}^p$. To remind the reader of that fact, we added the label '$at$' on the corresponding quantities.

Expressed in terms of the isospin coefficients, the effective contributions correspond to
\begin{eqnarray}
\frac{G_F}{\sqrt{2}}A\tilde C_S&=&AC_S^{(0)}-(Z_n-Z_p)C_S^{(1)}\quad\mbox{and}\\
\label{eq::CPTisospin}
\frac{G_F}{\sqrt{2}}\langle\sigma\rangle_{at}\tilde C_{P,T}^{at}&=&\langle\sigma\rangle_{at}C_{P,T}^{(0)}-(\langle\sigma_n\rangle_{at}-\langle\sigma_p\rangle_{at})C_{P,T}^{(1)}\,.
\end{eqnarray}
Note again that the coefficient of the triplet contribution is neither atom-independent nor generally small in the latter case; for example, $\langle\boldsymbol{\sigma}_p\rangle_{Xe}\approx\langle \boldsymbol{\sigma}_n\rangle_{Xe}/3$ and $\langle \boldsymbol{\sigma}_p\rangle_{\rm Hg}\approx\langle \boldsymbol{\sigma}_n\rangle_{\rm Hg}/10$ \cite{Dzuba:2009kn}, implying  
$(\langle\sigma_n\rangle_{at}-\langle\sigma_p\rangle_{at})/\langle\sigma\rangle_{at}\sim1$
for the latter. Note furthermore that the coefficient for $C_P^{(1)}$ is sometimes mistakenly given as $(Z_n-Z_p)/A$.

\subsubsection{The EDM of paramagnetic systems}
The EDM of paramagnetic systems is dominated to very good approximation by the contributions from $d_e$ and $\tilde C_S$, as explained above. The presently most constraining measurement from this class is performed with ThO~\cite{Baron:2013eja}. Their result is given in terms of an angular frequency, corresponding to an energy shift, which can be parametrized as
\begin{equation}
\omega = 2\pi\left(\frac{W_d}{2}d_e+\frac{W_c}{2}\tilde C_S\right)\,,
\end{equation}
using the conventions from \cite{Jung:2013mg} for the atomic constants $W_{d,c}$. 
For these, we obtain $W_d=-(3.67\pm0.18)\times 10^{25}{\rm Hz}/e\,{\rm cm}$ from \cite{PhysRevA.78.010502,2013arXiv1308.0414S,Fleig:2014uaa} and $W_c=-(598\pm90)\,{\rm kHz}$ \cite{2013arXiv1308.0414S}. Note that the calculations for the former are consistent; we use the value given in \cite{Fleig:2014uaa} (corresponding to $E_{\rm eff}=75.6\,{\rm GV/cm}$), and enlarge only slightly the uncertainty to $5\%$ due to the rather large difference to the central value in \cite{2013arXiv1308.0414S}. Note furthermore the consistency within uncertainties between the explicit calculations and the analytical estimate in \cite{PhysRevA.85.029901}.

\begin{figure}[hbt]
\centering{
\includegraphics[width=0.48\textwidth]{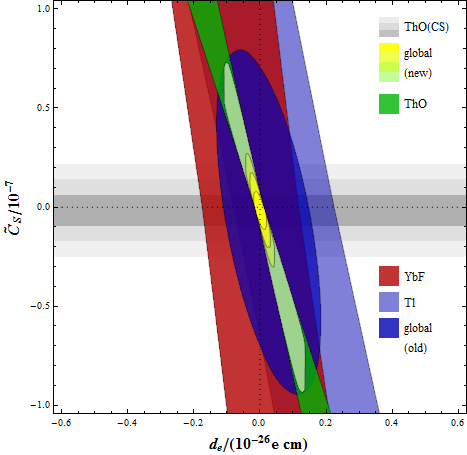}\quad\includegraphics[width=0.48\textwidth]{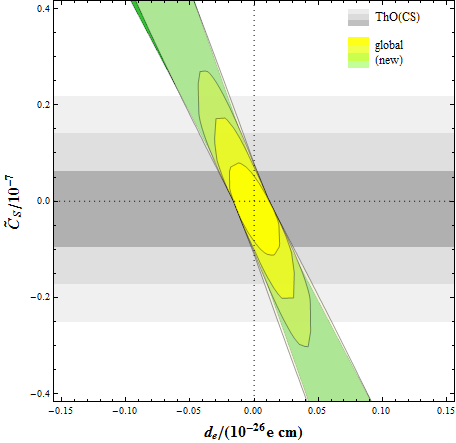}
}
\caption{The constraint for the electron EDM ($95\%$~CL) from the measurements in paramagnetic systems, see text. Left: global fit in comparison to the results from \cite{Jung:2013mg}. Right: zoom, showing only the ThO measurement \cite{Baron:2013eja} and the global fits. \label{fig::eEDMThO}}
\end{figure}

Since each measurement only constrains a combination of the two contributions, conservatively no constraint on the electron EDM alone can be obtained from any single measurement. The combination with previous measurements, performed with thallium (Tl) atoms and ytterbium fluoride (YbF) molecules \cite{Regan:2002ta,Hudson:2011zz} allows for a model-independent determination of the electron EDM, which improves significantly using in addition information from the mercury (Hg) system \cite{Jung:2013mg}, see Fig.~\ref{fig::eEDMThO}. However, in contrast to the situation in \cite{Jung:2013mg}, the Hg measurement does not provide a competitive bound on $\tilde C_S$ compared to the ThO one when setting $d_e$ to zero. Therefore, this procedure results in an  extremely conservative bound,
\begin{equation}\label{eq::eEDM}
|d_e|\leq 1.0\times 10^{-27}e\,{\rm cm}\quad(95\%~{\rm CL})\,,
\end{equation}
which is allowing for arbitrarily large cancellations between the different contributions and includes conservative estimates for the uncertainties of the various coefficients. In addition to this value, we obtain additional ones using assumptions on the maximal amount of fine-tuning: we restrict the contribution from $\tilde C_S$ alone not to exceed $n=1,2,3$ times the measured limit for ThO and use this as an additional constraint, thereby using effectively the ThO result twice. While this is clearly not as rigorous as the above limit, it is still more conservative than the common procedure to set the contribution from the electron-nucleon interaction simply to zero. This yields the inner solutions in Fig.~\ref{fig::eEDMThO}; the corresponding  upper limits for $d_e$ and $\tilde C_S$ are given in Table~\ref{tab::delimitnew}, together with the values quoted in \cite{Baron:2013eja}, which are obtained by setting the other contribution to zero and neglecting theory uncertainties. Note that, with a second competitive measurement, $d_e$ and $\tilde C_S$ can be extracted again without additional assumptions, see again \cite{Jung:2013mg}.
%
\begin{table}
{\centering{
\begin{tabular}{|l|l|l|}\hline
Input													& Limit for $|d_e|\,(95\%~{\rm CL})$		& Limit for $|\tilde C_S| (95\%~{\rm CL})$\\\hline\hline
Result w/o ThO \cite{Jung:2013mg}		 				& $1.4\times 10^{-27}e\,{\rm cm}$			& $7\times10^{-8}$\\\hline\hline
Including ThO, $\tilde C_S$ bounded by Hg				& $1.0\times 10^{-27}e\,{\rm cm}$			& $7\times10^{-8}$\\\hline
Including ThO, $\tilde C_S$ bounded by ThO ($n=3$)		& $0.35\times 10^{-27}e\,{\rm cm}$			& $2.3\times10^{-8}$\\
Including ThO, $\tilde C_S$ bounded by ThO ($n=2$)		& $0.25\times 10^{-27}e\,{\rm cm}$			& $1.6\times10^{-8}$\\
Including ThO, $\tilde C_S$ bounded by ThO ($n=1$)		& $0.16\times 10^{-27}e\,{\rm cm}$			& $0.8\times10^{-8}$\\\hline\hline
ThO only, $\tilde C_S=0$, $90\%$~CL \cite{Baron:2013eja}& $0.089\times 10^{-27}e\,{\rm cm}^{\dagger,\ddagger}$	& $0.6\times 10^{-8,\ddagger}$\\\hline 
\end{tabular}
\caption{\label{tab::delimitnew} New limits on the electron EDM and $\tilde C_S$, including the measurement in the ThO system \cite{Baron:2013eja}, see text. ${}^\dagger$: Using $W_d$ from \cite{2013arXiv1308.0414S}. ${}^\ddagger$: Theory errors neglected.}
}}
\end{table}
In the phenomenological section below, we use all values presented in Table~\ref{tab::delimitnew}, in order to demonstrate the progress due to the new measurement and to compare the various upper limits. We consider $n=2$ already a conservative choice, since there is no dynamical relation between the two contributions, rendering large cancellations unlikely. Nevertheless, the necessity to introduce this kind of assumption demonstrates the importance of independent measurements in other systems, ideally with strongly differing values for the ratio $W_d/W_c$ like, \emph{e.g.}, rubidium.

\subsubsection{The EDM of diamagnetic atoms}
For diamagnetic atoms mainly finite-size effects of the nucleus determine the EDM. More specifically, its main source is the CP-odd nuclear Schiff moment\footnote{Note that the operator used in the corresponding calculations receives corrections, the precise form of which are under discussion \cite{Liu:2007zf,PhysRevA.77.014101}. These corrections are, however, suppressed by $1/Z$ and therefore not relevant for the heavy systems under consideration here. Furthermore, there are relativistic corrections at the level $(Z\alpha)^2$ \cite{Flambaum:2001gq,Flambaum:2012vz}, which are included in a subset of the calculations, only.} \cite{Schiff:1963zz}. Although contributions from the nucleon EDMs are present as well, it is dominated by $T,P$-odd nuclear forces. These are represented by the interference of CP-even and -odd pion-nucleon interactions, the latter of which depend on the CEDMs of the up and down quark and four-quark operators (see again \eg{} \cite{Ginges:2003qt}). All of the necessary calculations are very involved, and the wide range of results indicates that the related theoretical uncertainties  are large; for recent discussions see,\eg, \cite{Ellis:2011hp,Engel:2013lsa}.

The first step, namely relating the atom EDM to the Schiff moment, is parametrized as
\begin{equation}\label{eq::dAdia}
d_{at}^{\rm dia}(S) = 10^{-17}e\,{\rm cm}\times C_{\rm Schiff}^{at}\times\frac{S}{e\,\mathrm{fm}^3}\,,
\end{equation}
with the constant $C_{\rm Schiff}^{at}$ being the result of multi-particle computations, modeling the electron-nucleon configurations in the corresponding atom. Due to the recent measurement in \cite{Griffith:2009zz}, the interest in these calculations has increased especially for Hg, leading to two recent results \cite{Latha:2009nq,Dzuba:2009kn},\footnote{Note that we disagree with the statement in \cite{Ellis:2011hp} that the sign of one calculation were incorrect.} from which we infer
\begin{equation}\label{eq::CSchiff}
C_{\rm Schiff}^{\rm Hg} = -2.6\pm 0.5\,\,\mbox{\cite{Dzuba:2009kn,Dilipprivcomm}}\,,
\end{equation}
which is now in agreement with the updated value of \cite{Latha:2009nq} (the preliminary result reads $C_{\rm Schiff}^{\rm Hg}=-2.46$ \cite{Dilipprivcomm}), strengthening the confidence in these calculations. The value also agrees with the earlier calculation \cite{Dzuba:2002kg} and is reasonably close to an old estimate \cite{Flambaum:1985ty}. 

In the next step, the Schiff moment is related to the CP-odd and -even $\pi NN$ coupling constants \cite{Barton:1961eg}, parametrized as \cite{Ban:2010ea} (note the different sign conventions for these constants used in the literature) 
\begin{equation}
S = g_{\pi NN}\left[(a_0+b)\,\bar{g}_{\pi NN}^{(0)}+a_1\,\bar{g}_{\pi NN}^{(1)}+(a_2-b)\,\bar{g}_{\pi NN}^{(2)}\right]\,.
\end{equation}
The isotensor coefficient is set to zero in the following, as its effect is suppressed by an additional factor of the mass difference of light quarks \cite{Pospelov:2001ys}. The CP-even coefficient is given by
$g_{\pi NN}=13.17\pm0.06$ \cite{Bugg:2004cm}, the  uncertainty of which is negligible in this context.
The corresponding nuclear calculations for mercury span a wide range and have in the case of $a_1$ also different signs in some of the calculations, 
see Table~\ref{tab::schiff}.
While in principle the calculations in \cite{Ban:2010ea} are more advanced than the previous ones, for mercury at some stage all the interactions used show problems, and the differences between the calculations are not well understood \cite{Ban:2010ea}; in absence of errors in one or several of the calculations, the problem might stem from the fact that mercury is a soft nucleus \cite{Ban:2010ea}.
\begin{table}
\centering{
\begin{tabular}{|l|c|c|c|}\hline
Ref. & $a_0\left[e\,{\rm{fm}}^3\right]$ & $a_1\left[e\,{\mathrm fm}^3\right]$ & $b\left[e\,{\rm fm}^3\right]$\\\hline
\cite{Dmitriev:2004fk} & $0.00004$ 
& 0.055 & --\\
\cite{deJesus:2005nb} & 0.010 (0.002-0.010) & 0.074 (0.057 - 0.090) & --\\
\cite{Ban:2010ea}${}^\dagger$ & (0.009 - 0.041) & (-0.027 - 0.005) & (0.002-0.013)\\\hline
\end{tabular}
}
\caption{\label{tab::schiff}Recent calculations for the coefficients relating the Schiff moment of mercury to the CP-violating pion-nucleon coupling constants. The values singled out in the second line are the ``preferred values'' quoted in the corresponding publication, the values in brackets show the range of values obtained with different Skyrme interactions, where available.
${}^\dagger$ Note that we do not agree with the authors of \cite{Ellis:2011hp} in that the SIII Skyrme interaction results in \cite{Ban:2010ea} were the most trustworthy of their calculations. In fact, it is shown in Ref.~\cite{deJesus:2005nb} that this interaction yields the worst results in reproducing the observables which can be used as experimental crosschecks, which is why the authors of \cite{Ban:2010ea} themselves regard it as critical.
}
\end{table}
We therefore estimate conservatively the following ranges:%
\begin{equation}
a_0 +b = (0.028\pm0.026)\,e\,{\rm fm}^3\quad\mbox{and}\quad a_1 = (0.032\pm0.059)\,e\,{\rm fm}^3\,,
\end{equation}
covering the full range of results shown in Table~\ref{tab::schiff}. We note that the possibility of vanishing $a_1$ implies that no constraint can be obtained conservatively on the corresponding isovector combination of CEDMs. Below we will show results for a representative value, in order to illustrate the potential of this observable, given a more reliable theoretical situation. Regarding the coefficient $a_0$, we point out that the tiny value obtained in \cite{Dmitriev:2004fk} might be the result of accidental cancellations, see the discussion in~\cite{deJesus:2005nb}. 
Finally, the parameter $b$ has so far only been calculated by one group; given the unclear situation, an additional independent calculation would be worthwhile. 

In the last step, the CP-odd $\pi NN$-couplings have to be related to the (C)EDMs of quarks. For this, typically a relation from the partially conserved axial current is used for the pion and QCD sum rules for the remaining nucleon matrix elements of quark currents. The main difficulty in this case is that for baryon sum rules in external fields the Borel transform does not exponentially suppress the contributions from excited states, leading to a large uncertainty. For the isovector coupling, this  can be improved by tuning an unphysical parameter to suppress these higher order terms, leading to \cite{Pospelov:2001ys}
\begin{equation}\label{eq::gb1}
\bar{g}_{\pi NN}^{(1)}(\tilde{d}) = \left(2_{-1}^{+4}\right)\times10^{-12}\frac{\tilde{d}_u-\tilde{d}_d}{10^{-26}{\rm cm}}\frac{|\langle\bar{q}q\rangle|}{(225~{\rm MeV})^3}\,.
\end{equation}
In the isoscalar sector a similar tuning is not possible, allowing for \cite{Pospelov:2001ys}
\begin{equation}\label{eq::gb0}
\bar{g}_{\pi NN}^{(0)} = (0.5\pm1.0)\times10^{-12}\frac{\tilde{d}_u+\tilde{d}_d}{10^{-26}{\rm cm}}\frac{|\langle\bar{q}q\rangle|}{(225~{\rm MeV})^3}\,,
\end{equation}
thereby also questioning the sensitivity to this combination of CEDMs.
An additional contribution to $g_{\pi NN}^{(1)}$ comes from four-quark operators, reading \cite{Shifman:1978zn,Anselm:1985cf,Demir:2003js,Ellis:2008zy}\footnote{Note that we correct here several typos with respect to the numerical evaluation in \cite{Demir:2003js}. Our result also slightly differs numerically from the one quoted in \cite{Ellis:2008zy}; we use in the evaluation Eq.~\eqref{eq::qc}, together with \cite{Beringer:1900zz} $\bar m(\mu_h)\equiv(m_u(\mu_h)+m_d(\mu_h))/2=4.7^{+0.9}_{-0.3}$~MeV, $f_\pi=92.4$~MeV and $m_\pi=137$~MeV.}
\begin{eqnarray}\label{eq::CqqME}
\bar{g}_{\pi NN}^{(1)}(C_{qq'}) &=& \frac{\langle \bar q q\rangle}{2f_\pi}\sum_{q=d,s,b}C_{qd}\left\langle N\left|\bar qq\right|N\right\rangle\\
&=& \frac{\langle \bar q q\rangle}{2f_\pi}\left(C_{dd}\frac{\sigma_{\pi N}}{m_u+m_d}+C_{sd}\kappa\frac{220~{\rm MeV}}{m_s}+C_{bd}\frac{2m_N}{3\tilde \beta m_b}(1-0.25\kappa)\right)\\
&=& -(6\pm3)\times 10^{-3}{\rm GeV}^3\left(0.6\frac{C_{dd}}{m_d}+3.3\kappa\frac{C_{sd}}{m_s}+(1-0.25\kappa)\frac{C_{bd}}{m_b}\right)\,,\label{eq::gpiNN4q}
\end{eqnarray}
where naive factorization has been used for the four-quark matrix elements.
Here $\kappa$ parametrizes the uncertainty in the strange quark content of the neutron, we use $\sigma_{\pi N}=\langle N|m_u\bar uu+m_d\bar dd|N\rangle=(59\pm7)$~MeV \cite{Alarcon:2011zs}, and $\tilde \beta=11-2n_l/3$ for $n_l$ light quarks \cite{Anselm:1985cf}. Recent lattice studies \cite{Young:2009zb,Toussaint:2009pz,Takeda:2010cw,Durr:2011mp} (see also \cite{Ohki:2009mt,MartinCamalich:2010fp,Bali:2011ks,Dinter:2012tt,Semke:2012gs}) indicate a smaller value for $\kappa$ than assumed previously (see \eg{} \cite{Borasoy:1996bx} and references therein), thereby reducing the influence of the strange quark on EDMs. However, while agreeing on a smaller order of magnitude, the range implied by these calculations is still relatively large. 
We combine them to arrive at
\begin{equation}\label{eq::kappa}
\kappa \equiv\frac{\langle N|m_s\bar{s}s|N\rangle}{220~{\rm MeV}}= 0.22\pm0.02\pm0.10\,,
\end{equation}
where we again chose a conservative range for the central value, reflected by the second uncertainty, while the first one is of statistical origin. 
However, as for the neutron, the four-quark contributions are subleading in \thdmws s, see the discussion in Sec.~\ref{sec::edmsin2hdms}. 

The Schiff moment receives contributions from the nucleon EDMs as well. While this contribution is not expected to be dominant, the resulting constraint for the neutron EDM is actually of the same order like the one from the dedicated experiments; using \eg{} Eq.~\eqref{eq::dAdia}, the range for $C_{\rm Schiff}^{\rm Hg}$ given in Eq.~\eqref{eq::CSchiff} and the expression $S(d_n)=1.9~{\rm fm}^2d_n$~\cite{Dmitriev:2003sc} (for simplicity with its central value),  we obtain $|d_n|\leq 7.8\times 10^{-26}e\,{\rm cm}$,  which is only about a factor of two weaker than the present direct limit \cite{Baker:2006ts}. However, there is no way to combine these limits, therefore we just consider the latter in the following.

Additional sources from electron-nucleon interactions and the electron EDM are present. Regarding the latter, the value usually used in the literature for mercury reads $d_{\rm Hg}(d_e)=1.16\times10^{-2}d_e$ \cite{1402-4896-36-3-011}. The corresponding calculation, however, shows a high sensitivity to higher order effects; the ``corrections'' to a previous estimate \cite{Flambaum:1985gx} amount to $\sim200\%$ and change the sign. The authors point out the sensitivity to correlation effects (which have been found to be large for mercury for the coefficients discussed above), making a new calculation mandatory. 
In light of this situation we do not see a way to extract a meaningful upper limit on the electron EDM from mercury until the theoretical situation improves. However, even taking the central value quoted above, the bound would be weaker than the one from paramagnetic systems.

The electron-nucleon interactions are induced via the three operators in Eq.~(\ref{eq::HeN}). In this case, the $\tilde C_S$ contribution is suppressed, as to leading order its contribution from closed electron shells vanishes; generically this leads to a dominance of the term proportional to $\tilde C_T$, if it is present. However, for the \thdmws s considered here, only the scalar-pseudoscalar operators are present. 
The contributions proportional to $\tilde C_P$ are often neglected, as its coefficient is one order of magnitude
smaller than the one of $\tilde C_S$, even in this case, due to the suppression by the nucleon mass. However, expressing the corresponding matrix elements in terms of coefficients of the four-fermion ones shows basically the opposite behaviour, rendering the sensitivity to fundamental parameters similar. All types of contributions are relevant in some part of parameter space \cite{Ellis:2008zy}. 

Given the large theoretical uncertainties in the contributions to the Schiff moment, the constraints on the electron-nucleon interaction might be the most important one at the moment. The coefficients in the relation $d_{\rm Hg}(\tilde C_{S,P})$ are obtained again in atomic calculations; usually only the coefficient of the tensor operator is calculated and approximate analytic relations are used to obtain the others\footnote{Note the different conventions for $d_{at}^{T,P}$ in different publications, \eg{} $\mathbf{d}_{at}^{T,P}=\langle\boldsymbol{\sigma}\rangle d_{at}^{P,T}$ versus $\mathbf{d}_{at}^{T,P}=\mathbf{I}/I d_{\rm at}^{T,P}$.} \cite{Flambaum:1985gx,Kozlov:1988qn,Ginges:2003qt,Dzuba:2009kn}:
\begin{eqnarray}
\label{eq::CSCT}\tilde C_S\frac{\mathbf{I}}{I}&\leftrightarrow&1.9\times10^3\left(1+0.3\,Z^2\alpha^2\right)^{-1}A^{-2/3}\mu^{-1}\times \tilde C_T^{at}\langle\boldsymbol{\sigma}\rangle\quad{\rm and}\\
\label{eq::CPCT}\tilde C_P^{N} &\leftrightarrow& 3.8\times10^3\frac{A^{1/3}}{Z}\tilde C_T^N\,,
\end{eqnarray}
where $\mu$ denotes the magnetic moment of the nucleus in terms of nuclear magnetons $\mu_N$. 
We expect the uncertainty for these relations to be relatively small, as also indicated by a recent explicit calculation for a variety of atoms \cite{Dzuba:2009kn}, which is why we neglect it in the following.

For the tensor coefficient, parametrized by
\begin{equation}
\mathbf{d}_{\rm Hg}(\tilde{C}_T) = C_{C_T}^{\rm Hg}\times10^{-20}\tilde{C}_T^{\rm Hg}\langle\boldsymbol{\sigma}\rangle e\,{\rm cm}\,,
\end{equation}
recent results read $C_{C_T}^{\rm Hg}=-5.1$ \cite{Dzuba:2009kn} and $C_{C_T}^{\rm Hg}=-4.3$ \cite{Latha:2009nq}.
Thus, using Eqs.~\eqref{eq::CSCT} and~\eqref{eq::CPCT}, we obtain
\begin{equation}
d_{\rm Hg}(\tilde C_S,\tilde C_P) = (1.0\pm0.1)(-4.7\, \tilde C_S+0.56\, \tilde C_P)\times10^{-22}e\,{\rm cm}\,,
\end{equation}
where we used $\mu_{\rm Hg}=0.506\,\mu_N$\footnote{Source: WebElements (\texttt http://www.webelements.com/)} and $\langle\boldsymbol{\sigma}\rangle=-1/3\,\mathbf{I}/I$, the estimate from a simple shell model for the nucleus, and the common convention $\mathbf{d}=d\mathbf{I}/I$.

The next step is to relate the coefficients $\tilde C_{S,P}$ to the effective operators discussed above. The contributing operators are four-fermion operators with electrons and light quarks, and an effective electron-2-gluon vertex from integrating out the heavy quarks. Again neglecting the up-type quark contributions, they can be parametrized as follows \cite{Shifman:1978zn,Anselm:1985cf,Demir:2003js,Ellis:2008zy}: 
\begin{eqnarray}\label{eq::CS}
\tilde C_S &=&\left(C_{de}\frac{\sigma_{\pi N}}{m_u+m_d}+C_{se}\kappa\frac{220~{\rm MeV}}{m_s}+C_{be}\frac{2m_N}{3\tilde \beta m_b}(1-0.25\kappa)\right)\\
&=&
\left(0.040\frac{\tilde C_{de}}{m_d}+0.220\,\kappa\frac{\tilde C_{se}}{m_s}+0.070(1-0.25\kappa)\frac{\tilde C_{be}}{m_b}\right){\rm GeV}\,,
\end{eqnarray}
where the same matrix element appears as for $C_{qq'}$, see Eq.~\eqref{eq::CqqME}.
The missing ingredients are the expressions for $\tilde C_P$ in terms of the coefficients of four-fermion operators. We use the estimates for the isospin coefficients (cf. Eqs.~\eqref{eq::CXisospin} and~\eqref{eq::CPTisospin}) \cite{Anselm:1985cf,Ellis:2008zy}
\begin{eqnarray}
\tilde C_P^{(0)}&\simeq&-0.375~{\rm GeV}\sum_{q=s,b}\frac{\tilde C_{eq}}{m_q}\,\qquad\mbox{and}\nonumber\\
\tilde C_P^{(1)}&\simeq& -0.806~{\rm GeV}\frac{\tilde C_{ed}}{m_d}-0.181~{\rm GeV}\sum_{q=s,b}\frac{\tilde C_{eq}}{m_q}\,,
\end{eqnarray}
again neglecting up-type quark contributions.

Finally, from these considerations we obtain the following result for mercury:
\begin{eqnarray}\label{eq::HgEDM}
d_{\rm Hg} &=& \left\{%
-(1.0\pm0.2)\left(%
(1.0\pm0.9)\,\bar{g}_{\pi NN}^{(0)}+1.1\,(1.0\pm1.8)\,\bar{g}_{\pi NN}^{(1)}\right)%
\right.\nonumber\\
&&%
+\left.(1.0\pm0.1)\times 10^{-5}\,\left[-4.7\,\tilde C_S+0.49\,\tilde C_P\right]\right\}\times 10^{-17}~e\,{\rm cm}\,,
\end{eqnarray}
with the expressions for $\bar{g}_{\pi NN}^{(1,0)}$ given in Eqs.~\eqref{eq::gb1}, \eqref{eq::gb0} and \eqref{eq::gpiNN4q}.

The possible vanishing of the coefficients of the isoscalar and -vector CEDM contributions implies that conservatively no bound can be obtained for them. Usually these contributions are assumed to be the dominant ones in this system, underlining the importance of theoretical developments to clarify the situation. Below, we will show the limits that would result for the central values in Eq.~(\ref{eq::HgEDM}), however only for illustration purposes.

\subsection{Renormalization of the effective operators\label{sec::RGErunning}}
To connect the relevant Wilson coefficients at the hadronic scale with the short-distance calculation at the electroweak one, the renormalization group running has to be taken into account. In general, QCD effects tend to reduce the value of the different coefficients (see \eg{} \cite{Degrassi:2005zd}),\footnote{Note that the anomalous dimensions of the operators $\mathcal{O}_W$ and $\mathcal{O}_f^\gamma$ have been used with the wrong sign in several publications in the past.} apart from the four-quark one \cite{An:2009zh,Hisano:2012cc}. We neglect its mixing into the CEDM because of its smallness; however, we take its enhancement into account in the estimate below. As pointed out in \cite{Degrassi:2005zd}, the mixing of the CEDM- into the EDM operators constitutes a large effect. 
On the other hand, we consider the NLO running of minor importance at present, given the large uncertainties in the hadronic matrix elements. Furthermore, the mixing of the Weinberg operator into the CEDM ones is of higher order in $\alpha_s$ and therefore neglected in the following. For models, in which this is not the case, the operator mixing has recently been discussed in~\cite{Hisano:2012cc}. Denoting $\mathbf{C}=(d_f^\gamma/2,d_f^C/2,C_W)$ (see Eq.~(\ref{eq::Leff})), this results in the following leading order expressions \cite{Dai:1989yh,Braaten:1990gq,Shifman:1976de,Degrassi:2005zd}:
\begin{eqnarray}\label{eq::RGE}
\frac{d\mathbf{C}_q(\mu)}{d\log\mu}&=&\frac{\alpha_s(\mu)}{4\pi}\left(\gamma^{(0)}_q\right)^T\mathbf{C}_q(\mu)\,,\quad\mbox{with}\\
\gamma^{(0)}_q &\equiv&\left(%
\begin{array}{c c c}%
\gamma_\gamma & 0 & 0\\%
\gamma_{C\gamma,q} & \gamma_C & 0\\%
0 & 0 & \gamma_W%
\end{array}%
\right)\nonumber\\ &=& \left(%
\begin{array}{c c c}
2C_F & 0 & 0\\
8C_F q_q & 10C_F-4N_C & 0\\
0 & 0 & N_C+2n_f
\end{array}%
\right)\,,
\end{eqnarray}
where $\beta_0=(11N_C-2n_f)/3$, $N_C=3$, $C_F=4/3$, $n_f$ denotes the number of active flavours and  $q_q=2/3,-1/3$ the charge for up- and down-quarks, respectively.

As we expect the Higgs masses to be of the order of $m_t$ (as is the mass of the already observed scalar), we choose $\mu_{tH}\sim m_t$ as the common matching scale where top quark and scalars are integrated out. We use the solution to Eq.~(\ref{eq::RGE}) to scale down to $\mu\sim m_b$, where in addition the beauty quark is integrated out, thereby matching $\mathcal{O}_b^C$ onto $\mathcal{O}_W$. 
The matching condition reads \cite{Braaten:1990gq,Boyd:1990bx}
\begin{equation}\label{eq::matching}
C_W(\mu_b^-)=C_W(\mu_b^+)+\frac{g_s^3}{8\pi^2m_b}\frac{d_b^C(\mu_b)}{2}\,,
\end{equation}
where $\mu_b^{+}(\mu_b^-)$ refers to the same scale $\mu_b$, but in the $n_f=5(4)$ theory, respectively. 
We emphasize that this matching, together with the larger anomalous dimension of the Weinberg operator, implies a relative enhancement of the contribution involving charged-scalar exchange compared to the one involving neutral scalars, as the suppression from the running is weaker for the CEDM contribution.
When going to the $n_f=3$ theory, the charm contribution to $C_W$ becomes local, which is however severely suppressed because of $m_c\ll m_t$, and therefore neglected.
The solution of Eq.~(\ref{eq::RGE}) reads
\begin{eqnarray}
\frac{d_q^\gamma(\mu_h)}{2} &=& \eta^{\kappa_\gamma}\frac{d_q^\gamma(\mu_{tH})}{2}+\frac{\gamma_{C\gamma}}{\gamma_\gamma-\gamma_C}\left(\eta^{\kappa_\gamma}-\eta^{\kappa_{C}}\right)\frac{d_q^C(\mu_{tH})}{2}\,,\label{eq::RGEdgamma}\\
\frac{d_q^C(\mu_h)}{2} &=& \eta^{\kappa_C}_{c-h}\eta^{\kappa_C}_{b-c}\eta^{\kappa_C}_{t-b}\frac{d_q^C(\mu_{tH})}{2}\,,\quad\mbox{and}\label{eq::runningC}\\
C_W(\mu_h) &=& \eta_{c-h}^{\kappa_W}\eta_{b-c}^{\kappa_W}\left(\eta^{\kappa_W}_{t-b}C_W(\mu_{tH})+\eta^{\kappa_C}_{t-b}\frac{g_s^3(\mu_b)}{8\pi^2m_b}\frac{d_b^C(\mu_{tH})}{2}\right)\,,\label{eq::runningW}
\end{eqnarray}
where we introduced $\eta_{i-j}=\alpha_s(\mu_i)/\alpha_s(\mu_j)$, $\eta=\eta_{t-h}$, and $\kappa_i=\gamma_i/(2\beta_0)$. For the sake of simplicity, Eq.~(\ref{eq::RGEdgamma}) is displayed for constant $n_f$ throughout the integration, but its change is taken into account in the numerical analysis.

Finally, regarding the Wilson coefficients of the semileptonic four-fermion operators, we note that they scale like the quark masses, therefore the combinations $C_{qe}/m_q$ and $C_{eq}/m_q$ are scale-independent.

\section{Experimental status\label{sec::exp}}
At present, the limits most sensitive on the various sources discussed above stem from searches for EDMs of Tl~\cite{Regan:2002ta}, YbF~\cite{Hudson:2011zz}, ThO~\cite{Baron:2013eja}, Hg~\cite{Griffith:2009zz} (see also \cite{2013PhRvA..87a2102S} for a more detailed discussion) and the neutron~\cite{Baker:2006ts}, see Table~\ref{tab::ExpEDM}. The physical origin of their EDMs is quite different, making them complementary sources of information. Although these limits have different orders of magnitude, their different dependences on the fundamental parameters of the theory actually lead to similar sensitivities.


\begin{table}
\centering{
\begin{tabular}{|c|c|c|}\hline
System & Present limit ($|d|/(e~{\rm cm})@95\%$~CL) & Expected limit (ST/MT)\\\hline
$n$ & $3.3\times 10^{-26}{}^{\dagger}$ \cite{Baker:2006ts} & $\mathcal{O}(10^{-27}/10^{-28})$\cite{vanderGrinten:2009zz,SNSnEDM,Altarev:2009zz,Altarev:2012uy,Masuda20121347,Serebrov2009263}\\\hline
$e$ & $\leq1.0\times10^{-27}$\,, see Table~\ref{tab::delimitnew} & $\mathcal{O}(10^{-29}/10^{-31})$ \cite{Amini:2007ku,2004APS..DMP.P1056K,Weissetal,Sakemi:2011zz,PhysRevX.2.041009,1367-2630-15-5-053034,Vutha:2009ux}\\\hline
$\phantom{}^{199}$Hg  & $3.1\times 10^{-29}$ \cite{Griffith:2009zz}    & $\leq 1\times 10^{-29}$ \cite{HgExp}\\
$\phantom{}^{129}$Xe  & $6.6\times 10^{-27}{}^\ddagger$ \cite{PhysRevLett.86.22}  & $\mathcal{O}(10^{-30})$\cite{Gemmel:2009pu,XenonExp}\\\hline
\end{tabular}
\caption{\label{tab::ExpEDM}Present limits on absolute values of EDMs at $95\%$~CL for the most sensitive systems, together with short term (ST) and mid term (MT) expected sensitivities.
${}^{\dagger}$:~Converted to $95\%$~CL, in the publication given at $90\%$~CL. 
${}^{\ddagger}$: Given in the paper as $(+0.7\pm 3.3\pm0.1)\times10^{-27}e\,{\rm cm}$.}
}
\end{table}

Several developments allow to expect significantly improved bounds or a non-zero measurement in the near future, see also \eg{} \cite{Ginges:2003qt,Pospelov:2005pr,Raidal:2008jk,Fukuyama:2012np,Engel:2013lsa}. The first option is to reduce the uncertainties within the established methods, but in the longer term techniques exploiting different features like octupole deformation hold the promise of qualitatively improving the sensitivities further. Regarding octupole deformation, important experimental progress has been reported recently in \cite{Gaffney:2013xx}. 

There are several experiments for the neutron EDM planned and running or under construction (see \cite{Serebrov:2013tba} for a recent result and again Table~\ref{tab::ExpEDM}) using different techniques to obtain higher neutron densities to achieve an up to two orders of magnitude improved bound. 

Regarding the electron EDM and the electron-nucleon coefficient $\tilde C_S$, with the experiments for thallium completely dominated by their systematic uncertainties, significant advancement seems difficult within this system. An improvement, again up to two orders of magnitude, might come instead from the cesium, rubidium and francium systems \cite{Amini:2007ku,2004APS..DMP.P1056K,Weissetal,Sakemi:2011zz,PhysRevX.2.041009}, which can be stored to obtain longer oscillation times. The expected limits correspond to probing the electron EDM down to  $\lesssim 10^{-29}e~{\rm cm}$ in the midterm future (2-3 years), and even $10^{-31}e\,{\rm cm}$ has been envisaged for the farther future in~
\cite{Sakemi:2011zz}. 
Further measurements with YbF are expected to strengthen the present limit in the short term~\cite{Hudson:2011zz} and various experiments are underway to gain sensitivity down to $\sim10^{-30}e\,{\rm cm}$ or further~\cite{Vutha:2009ux,1367-2630-15-5-053034} (see \eg{} \cite{Fukuyama:2012np,Engel:2013lsa} for a more complete list). 
A key technique is the rejection of systematic errors by using the so-called $\Omega$-doublet structure of a subset of paramagnetic molecules (characterized by two very closely lying states of opposite parity, leading to an extremely high polarizability), as demonstrated in the recent experiments \cite{2013PhRvA..87e2130E} -- so far obtaining a less stringent limit than the one from YbF -- and \cite{Baron:2013eja}.
 
In the context of the analysis in \cite{Jung:2013mg}, the expected presence of several measurements with similar sensitivities will allow to improve model-independently the limits on the electron EDM as well as the constant $\tilde C_S$, taking into account possible cancellations and at the same time removing the input from the Hg system and assumptions on fine-tuning.

In the future, trapped molecular ions might also be used as sensitive probes for EDMs, however, at the moment there are still severe experimental and theoretical challenges to overcome. Finally, also solid state systems are being explored as sensitive probes for the electron EDM \cite{Lamoreaux:2001hb,0038-5670-11-3-A11}. While again some experimental as well as theoretical progress is necessary before competitive results can be achieved, recent results show the progress in this field \cite{Eckel:2012aw}.

For diamagnetic systems, apart from some improvement from the Hg system itself~\cite{Griffith:2009zz,HgExp}, significant improvement is aimed at using xenon (${}^{129}{\rm Xe}$), for which the theoretical treatment is similar to the one described above. Further progress is expected with different enhancement mechanisms like intrinsic octupole moments and, related to that, closely neighboured parity doublets, which allow for large enhancement factors for the corresponding Schiff moments. Prominent examples are radium and radon; however, the calculation of the corresponding matrix elements is more complicated, making again theoretical uncertainties a critical issue. Furthermore, also diamagnetic molecules are under investigation. A first measurement exists in the TlF system \cite{Cho:1989hd}, but the planned experiments are expected to improve greatly on the present limits, see again \eg{} \cite{Fukuyama:2012np} for a recent list of experiments.  Generally, due to the various possible contributions to the EDMs, measurements in different diamagnetic systems are even more necessary to disentangle the sources and potentially differentiate between NP models. Ultimately, this could be done in an analysis similar to the one in paramagnetic systems \cite{Jung:2013mg}, but for that a lot more information than presently available is necessary. 

Finally, new techniques are being used for measuring the EDMs of charged particles directly by using a storage ring \cite{Garwin:1959xx,Semertzidis:1998sp,Khriplovich:1998zq,Farley:2003wt}, \eg{} for muons, where the present limit stems from a storage ring experiment already \cite{Bennett:2008dy}, the proton \cite{Semertzidis:2011qv}, which is supposed to be tested down to $10^{-29}e$~cm, or the Deuterium nucleus, which has the advantage of being lightly bound and allowing thereby to circumvent the large uncertainties present \eg{} in the nuclear calculations for mercury. There are also proposals to use the technique for molecular ions, see \eg{} \cite{Kawall:2011zz}.

\section{EDMs in 2HDMs\label{sec::edmsin2hdms}}
We now address the model-dependent second step in relating EDMs to model parameters, \ie{} calculating the relevant effective coefficients in specific models. The model dependence is in some sense more severe in EDMs than for other observables, for the following reason: as generic one-loop contributions are excluded already, an additional mechanism is necessary to render them small. This implies that the usual power counting is not sufficient, but that this suppression mechanism has to be incorporated.
As a result, even if a NP model has a 2HDM as an intermediate effective theory, this does not necessarily imply that limits calculated at that level hold for the full theory, as can be seen \eg{} in many SUSY models.

In this section we limit the discussion to \thdmws s. While generally even for these, limits cannot be given model-independently, 
we hold the discussion as general as possible, and the results, while given in the parametrization of the \athdmws, can be easily transferred to other frameworks.

\subsection{Contributions to EDMs in 2HDMs}
\begin{figure}
\centering{
\begin{minipage}{4.7cm}\begin{center}
\includegraphics[height=3.9cm]{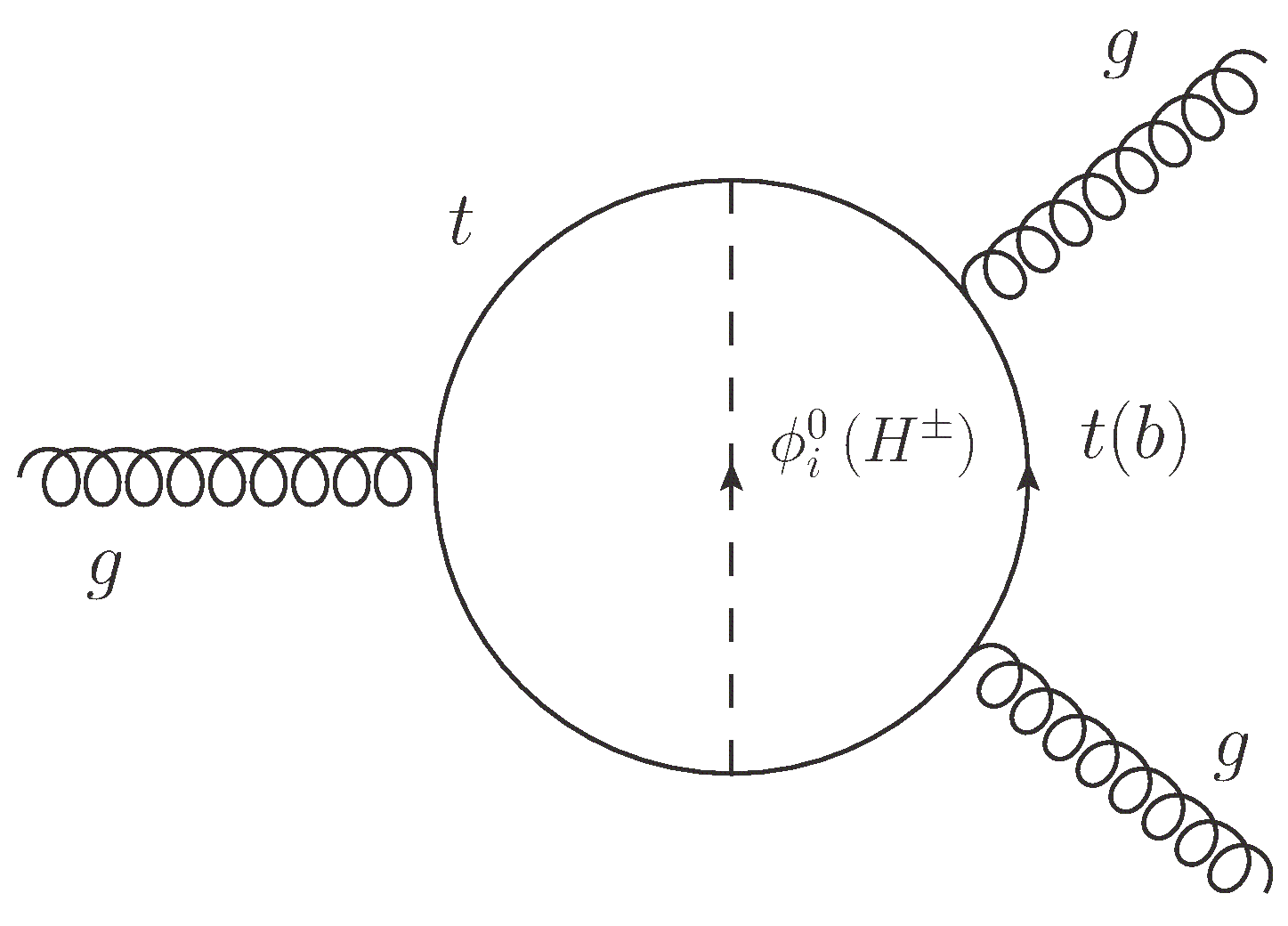}\\
(a)
\end{center}
\end{minipage}
\qquad\quad
\begin{minipage}{5.1cm}
\begin{center}
\includegraphics[height=3.9cm]{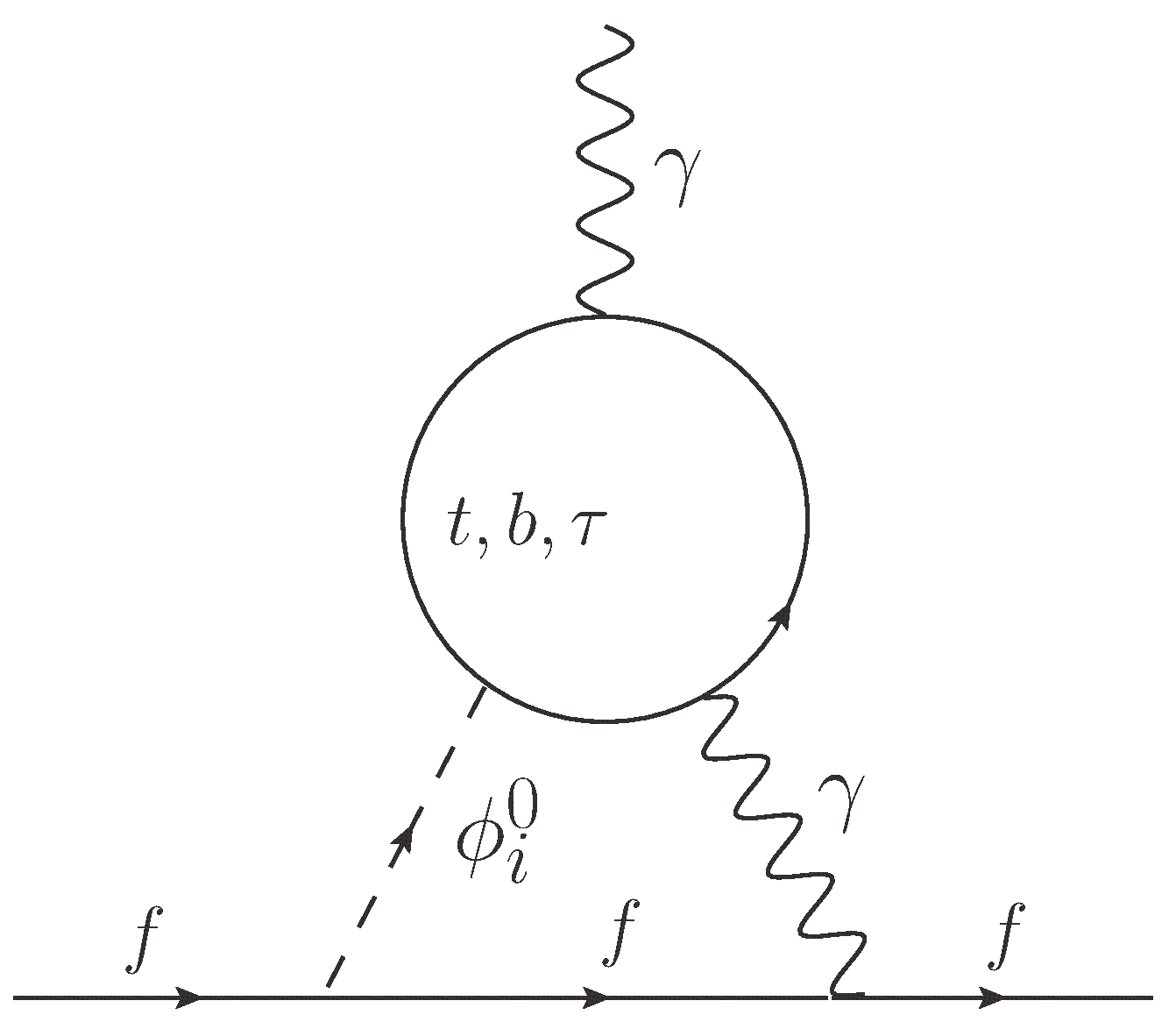}\\
(b)
\end{center}
\end{minipage}
\begin{minipage}{3.6cm}
\begin{center}
\vspace{1.0cm}
\includegraphics[height=2.1cm]{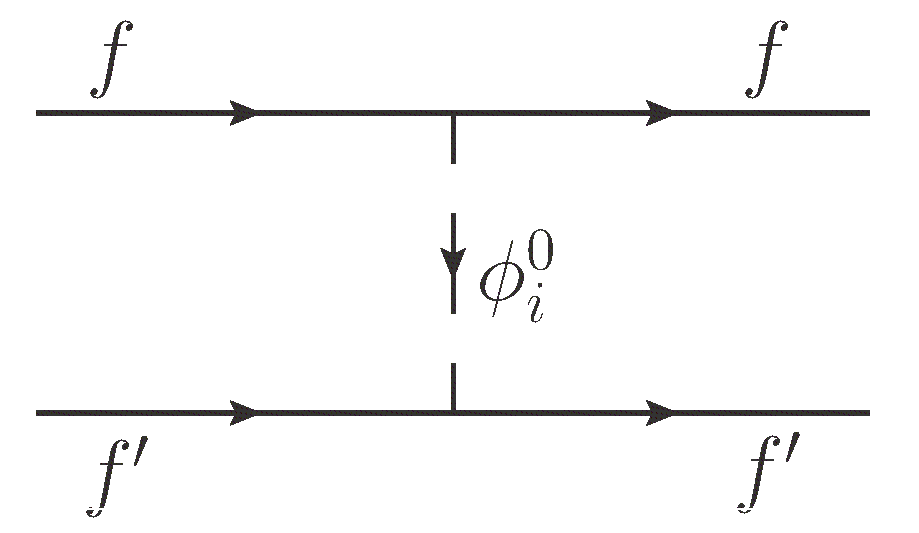}\\
\vspace{0.8cm}
(c)
\end{center}
\end{minipage}
\caption{\label{fig::diagrams}Classes of diagrams contributing to EDMs. (a) Contribution to the coefficient of the Weinberg operator. (b) Expample of a Barr-Zee diagram, contributing to all (C)EDMs. (c) Generic four-fermion contribution.}
}
\end{figure}

We start by listing the contributions to the different effective operators in Eq.~(\ref{eq::Leff}) within a \thdmws.
As most recent analyses have been done within
a SUSY framework, we will comment on the differences to the situation there when appropriate.
\begin{itemize}
\item Four-fermion operators: They induce the leading contributions in the SM~\cite{Pospelov:2005pr}, but there their effects remain extremely small. In \thdmws s, they are induced by CP-violating Higgs exchanges. While they can have contributions at tree level, in that case a further suppression by two light-fermion Yukawa couplings applies. 
If these are proportional to (or of the order of) the corresponding masses (as \eg{} in models with a $\mathcal{Z}_2$ symmetry, the \athdmws, MFV, Type III, $\ldots$), the ones with light fermions are suppressed to an acceptable level. The proportionality implies also that the induced coefficients divided by the corresponding masses are family-universal.
Those involving heavy fermions do not contribute directly, but induce higher-dimensional operators like $(\bar{f}f)\tilde{G}G$, again on an acceptable level, cf. Eqs.~\eqref{eq::Cbd} and~\eqref{eq::gpiNN4q}. 

There are two categories: CP-violating four-quark operators contribute to the nucleon EDM directly, or to the Schiff moments of nuclei by inducing CP-violating meson-nucleon interactions. As we will show below, for Higgs couplings of the order of the fermion masses, in the \thdm both contributions are subleading and can be neglected. 
In SUSY, they can receive contributions proportional to $\tan^3\beta$ from threshold corrections, rendering them more important there and even dominant for very large values of $\tan\beta$ \cite{Lebedev:2002ne,Demir:2003js}.

The second category consists of semileptonic operators. These induce CP-violating electron-nucleon couplings in atoms, as discussed in Sec.~\ref{sec::atomEDMs}. While in principle they are as suppressed as their four-quark equivalents, they receive very strong enhancement in heavy atoms due to the large number of nucleons and electrons, making their inclusion mandatory.
\item Weinberg operator: The contribution to this operator starts at the two-loop-level in \thdmws s, schematically shown in Fig.~\ref{fig::diagrams}(a). It is neither suppressed by small quark masses nor by small CKM elements, and therefore is expected to be large. However, for two reasons it is not completely dominating: first, the matrix element given in Eq.~(\ref{eq::dnCW}) is of the order of a light quark mass instead of a typical hadronic scale, and second the RGE running yields a strong suppression, see Eq.~(\ref{eq::RGE}).
As mentioned before, the second point is also the reason why,
contrary to naive expectations, the neutral Higgs contribution is generally suppressed compared to the charged one, cf. Sec.~\ref{sec::RGErunning}. In SUSY, the graphs discussed here are typically subleading, which is why they are often ignored.
\item (C)EDMs of light quarks: In the SM they vanish at the one- and even the two-loop level, leading to a tiny result \cite{Shabalin:1978rs}. In a general \thdmws, however, they can be generated at the one-loop level and are by far the leading contributions, which is why an additional mechanism for their suppression is necessary. An example are models with a $\mathcal{Z}_2$ symmetry, where these loops are CP-conserving like in the SM. In the \athdmws, but also more generally in models where the Higgs couplings are related to CKM-matrix elements and quark masses, the one-loop contribution for the light fermions is suppressed by at least one corresponding mass factor, together with factors like $m_U^2/M_{H^\pm}^2|V_{Ud}|^2$ or $m_D^2/M_{H^\pm}^2|V_{uD}|^2$ ($U=u,c,t,\,D=d,s,b$), rendering them one to two orders of magnitude smaller than the contributions discussed in the following. 
The reason is that the latter factors are circumvented in Barr-Zee(-type) diagrams \cite{Barr:1990vd,Gunion:1990iv,Chang:1990twa}, see Fig.~\ref{fig::diagrams}(b), which is why these two-loop contributions dominate in this class of models. The corresponding contributions are given later in this section.  They receive contributions from neutral scalars, discussed in the above papers, but also from charged ones. These contributions have been discussed partly \eg{} in \cite{BowserChao:1997bb,Chang:1999zw,Pilaftsis:1999td}, but to the best of our knowledge \eg{} 
the ones for the down-quark EDM with a top  and beauty quark in the loop are still missing. To construct a Barr-Zee diagram with a charged Higgs, a second charged current is necessary; therefore there are no contributions to the CEDMs from these graphs.

In SUSY, there are more one-loop contributions from loops with gauginos and sfermions, generally leading to strong bounds on the imaginary parts of the corresponding couplings. From the Higgs sector, the two-loop contributions again dominate, due to the arguments given above.
\item Electron EDM: The SM contribution to this is tiny, as for $m_\nu\to0$ it vanishes even on the three-loop level \cite{Pospelov:1991zt}. In \thdmws s, the one-loop contributions are real unless a neutrino coupling is involved, which is why the dominant contributions are again on the two-loop level, from Barr-Zee diagrams. In SUSY, already on the one-loop level sizable contributions appear from gaugino-slepton loops, therefore again the Higgs contributions do not receive that much attention. 
\end{itemize}
Because of the arguments given above, we will explicitly consider only the contributions stemming from the two-loop diagrams and the semileptonic four-fermion operators important for atoms and molecules.
It should be emphasized again that the limits obtained within \thdmws s are sensitive to the UV completion of the model, as their sensitivity to two-loop contributions already shows. Especially in SUSY there are usually large one-loop contributions dominating, which are not included here.\\

The contributions listed above are related to different sources of CP violation in \thdmws s: while \eg{} the charged Higgs contribution to the Weinberg operator stems only from CP violation in the Yukawa couplings of the model, diagrams involving neutral scalars in general receive contributions from the Higgs potential as well. Before  providing results for specific diagrams, we discuss the different classes of contributions, pointing out their general features.

\subsection{Charged Higgs contributions in \thdmws s}
The Lagrangian for charged Higgs exchange can for vanishing neutrino masses be parametrized as
\begin{equation}\label{eq::lagrangiancharged}
\mathcal L_Y^{H^\pm} \! =\! - \frac{\sqrt{2}}{v}\, H^+ \left\{  \bar{u}   \left[  V \varsigma_d M_d \mathcal P_R - \varsigma_u\, M_u^\dagger V \mathcal P_L  \right]   d\, +\,  \bar{\nu} \varsigma_l M_l \mathcal P_R l\right\} + \;\mathrm{h.c.} \, ,
\end{equation}
where $V$ is the CKM mixing matrix and the form reflects the fact that we will be mostly  concerned with the \athdmws, where $\varsigma_{u,d,l}$ are complex numbers of $\mathcal{O}(1)$; for a general \thdmws, they are arbitrary matrices and the dependences on the quark mass matrices are artificial, \ie{} just a possible normalization. 
Note that we consider the latter form simply as a phenomenological parametrization. For the contributions calculated explicitly  below, this implies only the generalization of the factors $\varsigma_{u,d,l}$. However, when these elements are indeed arbitrary, other contributions are possibly dominant, since especially the suppression  for the one-loop contributions to (C)EDMs explained above can be spoiled. If the scaling does approximately hold, the bounds obtained below are valid for the corresponding generalized couplings.

\subsection{\label{sec::neutralHiggs}Neutral Higgs contributions in \thdmws s}
The flavour-diagonal Higgs couplings are parametrized analogously as
\begin{equation}\label{eq::lagrangianneutral}
\mathcal L_Y^{\varphi_i^0} \! =\! -\frac{1}{v}\; \sum_{\varphi, f}\,  \varphi^0_i  \; \bar{f}\,y^{\varphi^0_i}_f\,  M_f \mathcal P_R  f\;  + \;\mathrm{h.c.} \, ,
\end{equation}
with the fields $\varphi_i^0=\{h,H,A\}$ denoting the neutral scalar mass eigenstates. Introducing the notation $F(f)$ for the species of a fermion, \eg{} $F(u)=F(c)=F(t)=u$, we write the fermion couplings to neutral scalars as
\begin{eqnarray}
y_{f}^{\varphi^0_i} &=& \cR_{i1} + (\cR_{i2} + i\,\cR_{i3})\,\left(\varsigma_{F(f)}\right)_{ff}\quad {\rm for}\quad F(f)=d,l\,,\quad{\rm and}\\
y_f^{\varphi^0_i} &=& \cR_{i1} + (\cR_{i2} -i\,\cR_{i3}) \,\left(\varsigma_{F(f)}^*\right)_{ff} \quad {\rm for}\quad F(f)=u,
\end{eqnarray}
to allow for the general form of $\varsigma_{u,d,l}$.
Here, $\cR$ is an orthogonal matrix defined by $\mathcal{M}^2_{\rm diag}=\mathcal{R} \mathcal{M}^2\mathcal{R}^T$, relating the mass eigenstates to the neutral scalar fields $S_i$ in the Higgs basis, where $\Phi_1^T= [ G^+,(v+S_1+iG^0)/\sqrt{2}]$ and $\Phi_2^T= [ H^+,(S_2+iS_3)/\sqrt{2} ]$,
and the fields $G^i$ are the Goldstone bosons absorbed by the gauge bosons: $\varphi_i^0=\cR_{ij}S_j$. In a general \thdmws, the  $\varsigma_{u,d,l}$ are the matrices introduced in Eq.~\eqref{eq::lagrangiancharged}, only the diagonal elements of which are relevant here.

The fact that these interactions involve three neutral bosons, two of which have unknown masses, and that the matrix $\cR$ depends on the scalar potential, which is largely unknown so far as well, renders these contributions very hard to deal with phenomenologically, even when a specific model like the \athdm is assumed. To avoid these difficulties, in the literature typically the dominance of the contribution from the lightest scalar is assumed; this is however problematic, as we will discuss below. While we will still apply this assumption occasionally to obtain indicative numbers for the neutral couplings, we can also include new information compared to earlier analyses: thanks to the huge amount of data collected recently by the LHC experiments and to lesser extend the Tevatron ones, we have some information already on the matrix $\cR$.
The collider data shows that the Higgs-like state discovered at the LHC couples to $W^+W^-$ and $ZZ$ with a strength close to the SM one;
assuming that it corresponds to the lightest neutral scalar $h$, one gets
$|\cR_{11}|> 0.80$ at 90\% CL \cite{Celis:2013rcs,Pich:2013vta}.
The orthogonality of $\cR$ implies then $\sqrt{|\cR_{21}|^2+|\cR_{31}|^2}< 0.60$ at 90\% CL.

We now return to the assumption of dominating contributions from the lightest Higgs. Since EDMs are T and therefore CP violating, the contributions from neutral scalars typically involve the combinations ${\rm Re}\left(y_f^{\varphi^0_i}\right){\rm Im}\left(y_{f'}^{\varphi^0_i}\right)$.  While the rotation matrix $\cR$ is unknown (to the extend discussed above), we do know that it is orthogonal. This property yields one central relation for these couplings ($\xi_{d,l} = 1,\, \xi_u = -1$):
\begin{equation}\label{eq::ycancellation}
\sum_i {\rm Re}\left(y_f^{\varphi^0_i}\right){\rm Im}\left(y_{f'}^{\varphi^0_i}\right)=\xi_{f'}\;{\rm Im}\left[(\varsigma_{F(f)}^*)_{ff}(\varsigma_{F(f')})_{f'f'}\right]\,.
\end{equation}
This sum is therefore independent of the scalar potential and obviously vanishes for $f=f'$, couplings $\varsigma_{f,f'}$ with identical phases (\eg{} real couplings, as for example present in $\mathcal{Z}_2$ models or MFV as defined in \cite{D'Ambrosio:2002ex}), and also for $F(f)=F(f')$ when the $\varsigma_{f,f'}$ are family-universal (as in the \athdmws). In the expressions below, the terms are weighted typically by some function of the neutral Higgs masses, making it most relevant for degenerate Higgs masses. However, the expression implies that all contributions stemming from CP violation in the potential involve mass differences of the neutral scalars, and that generally large cancellations can be expected in the neutral sector.

The precise form of the matrix $\mathcal{R}$ depends on the potential for the scalar fields; note that in general the mass eigenstates do not correspond to CP eigenstates. For a CP-invariant potential, specifically, the rotation takes the simple form
\begin{equation}\label{eq::RCPC}
\cR_{\rm CPC}=\left(
\begin{array}{c c c}
\cos\tilde \alpha & \sin\tilde \alpha & 0\\
-\sin\tilde \alpha & \cos\tilde \alpha & 0\\
0 & 0 & 1
\end{array}\right)\,,
\end{equation}
where $\tilde \alpha$ is often denoted $\alpha-\beta$ in $\mathcal{Z}_2$ models, leading to 
\begin{eqnarray}
{\rm Re}y_u^{\varphi^0_1}{\rm Im}y_u^{\varphi^0_1}&=&-{\rm Im}(\varsigma_u)\sin\tilde \alpha(\cos\tilde \alpha+{\rm Re}(\varsigma_u)\sin\tilde \alpha)\,,\nonumber\\
{\rm Re}y_u^{\varphi^0_2}{\rm Im}y_u^{\varphi^0_2}&=&-{\rm Im}(\varsigma_u)\cos\tilde \alpha(-\sin\tilde \alpha+{\rm Re}(\varsigma_u)\cos\tilde \alpha)\,,\nonumber\\
{\rm Re}y_u^{\varphi^0_3}{\rm Im}y_u^{\varphi^0_3}&=&{\rm Im}(\varsigma_u){\rm Re}(\varsigma_u)\,,
\end{eqnarray}
with similar expressions for the remaining combinations. 
Note that in this case all contributions vanish for real $\varsigma_{u,d}$, while in general mixing between the CP-odd and -even components can induce CP violation as well. 

The general argument above is strengthened by a second important observation, namely that the 
scalar mixing angles are not independent of their masses.
To be specific, let us consider the limit where the second scalar doublet $\Phi_2$ receives a very large mass and decouples from the low-energy effective theory. In this limit, the (SM-like) light Higgs has a mass $M^2_h\sim \cO(v^2)$, while all the other scalars become heavy and degenerate, {\it i.e.}, $M_H^2 = M_A^2 = M_{H^\pm}^2$ up to corrections of $\cO(v^2)$.
If the potential is CP symmetric, the mixing angle in Eq.~\eqref{eq::RCPC} vanishes in the decoupling limit: $\tan{\tilde \alpha}\sim \cO(v^2/M_{H^\pm}^2)$.
More generally, allowing for CP violation in the scalar potential, this limit yields the following form for the scalar mixing matrix:
\begin{equation}
\cR_{\rm dec} \; =\;\left(
\begin{array}{c c c}
1 & 0 & 0\\
0 & \cos{\theta_{\mathrm{CP}}} & -\sin{\theta_{\mathrm{CP}}}\\
0 & \sin{\theta_{\mathrm{CP}}} & \cos{\theta_{\mathrm{CP}}}
\end{array}\right)
\; +\; \cO(v^2/M_{H^\pm}^2)
\,,
\end{equation}
with some potential-dependent angle $\theta_{\mathrm{CP}}$ which vanishes if CP is conserved.\footnote{The exact relation is $\tan{(2\theta_{\mathrm{CP}})} = \mathrm{Im}(\lambda_5)/\mathrm{Re}(\lambda_5)$ with $\lambda_5$ one of the scalar potential parameters defined in \cite{Celis:2013rcs}.}
This implies ${\rm Im}y_f^{\varphi^0_1}=0$ and, therefore, the cancellation of Eq.~\eqref{eq::ycancellation} takes place only between $\varphi^0_2$ and $\varphi^0_3$, which in addition have equal masses in this limit. Thus, in the absence of complex Yukawa couplings, the sum of scalar contributions would vanish even with mass-dependent weight factors.
This fact is sometimes overlooked in the literature, leading to claims of non-vanishing contributions in the decoupling limit which are not correct in this context.

Together, these observations imply two strong statements:
\begin{enumerate}
\item For a vanishing right-hand side in Eq.~\eqref{eq::ycancellation}, EDM contributions from neutral scalars in \thdm vanish for small as well as very large mass differences. Therefore, generally large cancellations can be expected.
\item Even with the right-hand side present, the only contribution not suffering this suppression stems from the factors $\varsigma_{u,d,l}$ which determine also the charged Higgs interactions.
\end{enumerate}
In both cases, the approximation of simply taking the contribution from the lightest Higgs is not a good one; specifically, it is not conservative.

\subsection{The Aligned Two-Higgs-Doublet Model}
We are now prepared to proceed and give the expressions for the relevant coefficients within specific models. We do this exemplarily for the  \athdm \cite{Pich:2009sp,Jung:2010ik}. 
We discuss here only the constraints from EDMs; for  other  phenomenological constraints, see \cite{Jung:2010ab,Jung:2012vu,Celis:2012dk,Celis:2013rcs}.

In the \athdmws, the problem of FCNCs is circumvented by assuming at some scale $\Lambda_A$ alignment of the two Yukawa matrices present for each fermion species. The flavour-changing Higgs couplings are then determined by the CKM matrix and the three complex parameters $\varsigma_{u,d,l}$ mentioned above, constituting new sources for CP violation. The various models with $\mathcal{Z}_2$ symmetry appear as limiting (CP-conserving) cases of these parameters, see \cite{Pich:2009sp}. While renormalization induces some misalignment, the structure of the model prevents these effects from becoming sizable \cite{Pich:2009sp,Jung:2010ik,Braeuninger:2010td,Ferreira:2010xe}. Note, however, that the operators additionally generated by the misalignment are not relevant in this context, since they are not flavour-diagonal.

The resulting Yukawa couplings have the form given in Eqs.~\eqref{eq::lagrangiancharged} and~\eqref{eq::lagrangianneutral}, where now $\varsigma_{u,d,l}$ are complex numbers of $\mathcal{O}(1)$ instead of matrices. Specifically, as mentioned above, the right-hand side of Eq.~\eqref{eq::ycancellation} vanishes in this case for two fermions of the same electric charge. 

We now turn to calculating the expressions for the different classes of diagrams in the \athdmws, contributing to the effective coefficients in Eq.~(\ref{eq::Leff}). The phenomenological analysis of these expressions is postponed to the next section.

\subsection{Four-fermion operators}
For the four-fermion interactions, cf. Fig.~\ref{fig::diagrams}(c), we obtain for the \athdmws
\begin{equation}
C_{ff'}=\sqrt{2}G_F\sum_{i}\frac{m_fm_{f'}}{M_{\varphi_i^0}^2}{\rm Re}\left(y_f^{\varphi^0_i}\right){\rm Im}\left(y_{f'}^{\varphi^0_i}\right)\,.
\end{equation}
Because the neutral Higgs coupling $y_f^{\varphi^0_i}$ is identical for fermions of the same charge, the ratio $C_{ff'}/(m_fm_{f'})$ is rendered family-independent. 
As noted above, the electron-quark couplings are important for the EDMs of atoms and molecules.
An estimate of the contributions for the neutron from four-quark operators on the other hand reads
\begin{equation}
d_n^{4f}\sim 7\times 10^{-28}e\,{\rm cm} \sum_{i}{\rm Re}y_d^{\varphi^0_i}\,{\rm Im}y_d^{\varphi^0_i} \frac{(125~{\rm GeV})^2}{M_{\varphi_i^0}^2}\,,
\end{equation}
where we included the RGE enhancement by an approximate factor of five, cf. \cite{Hisano:2012cc}. Note that additionally the cancellation discussed above has to be considered.
This implies at most a moderate contribution, which is well below the two-loop contributions discussed later. Therefore, we neglect it in the following. An analogous statement holds for mercury.

\subsection{The Weinberg operator}
As mentioned before, the Weinberg operator is of special importance, as its contribution is neither suppressed by light quark masses nor by small CKM elements. Here we have calculated the different contributions in the \athdmws; our results agree with the results obtained in \cite{Weinberg:1989dx,Dicus:1989va} when translating them into the language of complex propagators used there.

\subsubsection{Charged Higgs contribution}
As described in Sec.~\ref{sec::RGErunning}, we perform the analysis of the charged Higgs contribution in an effective field theory framework \cite{Braaten:1990gq,Boyd:1990bx}, which simplifies the problem to the calculation of one-loop diagrams.  
The corresponding amputated diagram contains contributions from two operators; the correct coefficient can be read off from the Dirac structure $\gamma^\mu\fmslash{q}\gamma_5$, for which the additional contribution is absent \cite{Grinstein:1987vj}.
Our result reads
\begin{equation}\label{eq::C2}
\frac{d_b^C(\mu_{tH})}{2}=-\frac{G_F}{\sqrt{2}}\frac{1}{16\pi^2}|V_{tb}|^2m_b(\mu_{tH})\;\mbox{Im}(\varsigma_d\varsigma_u^*)\;x_{tH}\left(\frac{\log(x_{tH})}{(x_{tH}-1)^3}+\frac{x_{tH}-3}{2(x_{tH}-1)^2}\right)\,,
\end{equation}
where $x_{tH}=m_t^2/M_{H^\pm}^2$, which is to be used in Eq.~\eqref{eq::runningW} to obtain its contribution to $C_W$.
We have checked that this result agrees with \cite{Boyd:1990bx}, noting that their Lagrangian for charged Higgs exchange corresponds to ours for $n=2$ scalar doublets, $-Y_{12}/Y_{11}=\varsigma_u$ and $Y_{22}/Y_{21}=-\varsigma_d$.
Our common result in turn corresponds at the matching scale to the one obtained in \cite{Dicus:1989va}.

\subsubsection{Neutral Higgs contribution}
For the neutral Higgs contribution the full two-loop diagram has to be calculated, as for a top quark in the loop internally only heavy degrees of freedom appear. The calculation proceeds via the following steps: 
the three-gluon matrix element is obtained by 
using from every field-strength tensor only the part containing derivatives, and summing over all possible permutations,
leading to ($p_3=-p_1-p_2$)
\begin{eqnarray}
\left\langle\mathcal{O}\right\rangle\!\! &=&\!\!
-\frac{2}{3}f_{abc}C_W\,\epsilon_a^\mu(p_1)\epsilon_b^\nu(p_2)\epsilon_c^\rho(-p_1-p_2)
\left[(p_1-p_2)_\rho\,\epsilon_{\alpha\beta\mu\nu}+2\left(p_{1\,\nu}\,\epsilon_{\alpha\beta\mu\rho}+p_{2\,\mu}\,\epsilon_{\alpha\beta\nu\rho}\right)\right]p_1^\alpha p_2^\beta\,.
\end{eqnarray}
Here we ignored higher orders in $p_i^2/M^2$ ($M\in\{M_{H},m_t\}$) and used $\epsilon_a^\mu(p_1)\epsilon_{b\,\mu}(p_2)=\epsilon_b^\mu(p_1)\epsilon_{a\,\mu}(p_2)$ as well as $p_\mu\epsilon_a^\mu(p)=0$.

The other side of the matching condition is calculated by again summing over the different momentum configurations for the two-loop diagram, identifying the part proportional to the same Dirac structure in the corresponding expression, expanding carefully in the external momenta, and using the Feynman parametrization for the remaining integrals. The different integrals combine to give the result%
\footnote{Note that in principle the correct procedure for the $b$-quark contribution would be analogous as for the charged Higgs contribution, \ie{} integrating out the Higgs, running the resulting 4-quark operator down to $\mu\sim m_b$ and matching it on the Weinberg operator. This produces a potential enhancement from a smaller anomalous dimension. However, considering the enhancement for the charged Higgs, the resulting contribution would be at most on the level of the one from Barr-Zee diagrams discussed below. As their relative sign is unknown, it would therefore not improve the limit on ${\rm Im}(y_d^2)$ given later, which is why we use this simplified treatment.}
\begin{equation}
C_W(\mu_{tH}) = %
4g^3(\mu_{tH})\frac{\sqrt{2}G_F}{(4\pi)^4}\sum_{f=t,b}\sum_{i}{\rm Re}\left(y_f^{\varphi^0_i}\right){\rm Im}\left(y_f^{\varphi^0_i}\right)h(m_f,M_{\varphi_i^0})\,,
\end{equation}
which is again in agreement with \cite{Weinberg:1989dx} for the top-quark contribution, where however the $b$-quark one was neglected.
Here, $h(m,M)$ is defined by\footnote{The inner integral can be done analytically, 
simplifying the numerical analysis. Note the factor of 2 between the definition of $h(m,M)$ in \cite{Weinberg:1989dx} and \cite{Dicus:1989va}, the latter of which we are using here.}
\begin{equation}
h(m,M) = \frac{m^4}{4}\int_0^1dx\int_0^1du\frac{u^3x^3(1-x)}{[m^2x(1-ux)+M^2(1-u)(1-x)]^2}\,.
\end{equation}

Naively, observing the factor $m^4$ in the definition of $h(m,M_{\varphi_i^0})$, it seems unnecessary to include the beauty quark in the loop. However, the parametric integral diverges for $m\to 0$; for $M_{\varphi_i^0}\gg m$ the limit reads
\begin{equation}
h(m,M\gg m)=\frac{1}{4}\frac{m^2}{M^2}\left[\log\left(\frac{M^2}{m^2}\right)-\frac{3}{2}\right]\,,
\end{equation}
implying a much weaker suppression of the corresponding contribution, which might be compensated if $|\varsigma_d|\gg |\varsigma_u|$.

\subsection{Barr-Zee diagrams}

The diagrams for EDMs introduced by Barr and Zee \cite{Barr:1990vd} (and later generalized for the gluonic dipole moment \cite{Gunion:1990iv,Chang:1990twa}, see furthermore \cite{DeRujula:1990wy,Gunion:1990ce,Leigh:1990kf,Chang:1990sf,Kao:1992jv}) are proportional to the light quark mass, which at first sight leads to the conclusion that they should be tiny compared to the contribution from the Weinberg operator, which does not suffer this suppression. However, the following arguments show that their contributions are in fact comparable (cf. also \cite{Gunion:1990iv}):
\begin{itemize}
\item $e^3$ and $e \,g_s^3$ are of similar size at $\mu_{tH}$.
\item The anomalous dimension for the Weinberg operator is larger, implying a stronger suppression from the running.
\item The parametric integral of the Weinberg operator is smaller.
\item Finally, the matrix element of the Weinberg operator is very small
\cite{Demir:2002gg}, making it comparable to the mass of a light quark.
\end{itemize}
Therefore these contributions have to be taken into account. Which kind of diagram dominates depends in part on the method chosen to estimate the matrix elements, which we discussed in Sec.~\ref{sec::MIexpressions}.

\subsubsection{Neutral Higgs contribution}
In \cite{Barr:1990vd}, the neutral Higgs contributions are calculated for a quark and gauge bosons in the loop, while those with internal scalars are neglected. This contribution, however, is generally smaller than the others \cite{Kao:1992jv} and we will not discuss it here. As the paper is formulated for CP-violating Higgs propagators, the translation to our model parameters is not always trivial. Especially it is not universal; what is called $Z_2$ for example in \cite{Weinberg:1989dx,Barr:1990vd} changes for the type of diagram considered. 
Starting with the diagrams with the neutral Higgs between two fermions $f,f'$, cf. Fig.~\ref{fig::diagrams}(b), it reads
\begin{equation}\label{Eq::tanslationZ0y}
Z_{0n}=y_f^{\varphi^0_i}y_{f'}^{\varphi^0_i}\quad{\rm and}\quad \tilde{Z}_{0n}=y_f^{\varphi^0_i}y_{f'}^{\varphi^0_i*}\,,
\end{equation}
implying ${\rm Im}\,\tilde{Z}_{0n}=0$ for $f=f'$.
This relation implies the following contributions for the EDM/CEDM of a fermion $f$ with a fermion $f'$ in the loop via neutral Higgs exchange\footnote{Note that the correction for the sign for ${\rm Im}(Z_1)$ in the erratum of \cite{Barr:1990vd} applies to the whole paper.}, generalizing slightly the results of \cite{Barr:1990vd,Gunion:1990iv}:
\begin{eqnarray}\label{eq::EDMBZ}
\frac{d_{f}^{\gamma}(\mu_{tH};\varphi^0)_{\rm BZ}}{2} &=& -2\frac{\sqrt{2}G_F\alpha}{(4\pi)^3}m_f q_{f}\sum_{f'}\sum_{i}q_{f'}^2N_C^{f'}\left\{f\!\left(\frac{m_{f'}^2}{M_{\varphi_i^0}^2}\right)\left(2{\rm Re}y_f^{\varphi^0_i}{\rm Im}y_{f'}^{\varphi^0_i}\right)\right.\nonumber\\
&&\hskip 4.9cm +\;\left.g\!\left(\frac{m_{f'}^2}{M_{\varphi_i^0}^2}\right)\left(2{\rm Re}y_{f'}^{\varphi^0_i}{\rm Im}y_f^{\varphi^0_i}\right)\right\}\,,\quad{\rm and}\\
\label{eq::CEDMBZ}
\frac{d^{C}_q(\mu_{tH})_{\rm BZ}}{2} &=& -\frac{\sqrt{2}G_F\alpha_s}{(4\pi)^3}m_q\sum_{q'}\sum_{i}\left\{f\!\left(\frac{m_{q'}^2}{M_{\varphi_i^0}^2}\right)\left(2{\rm Re}y_q^{\varphi^0_i}{\rm Im}y_{q'}^{\varphi^0_i}\right)\right.\nonumber\\
&&\hskip 3.4cm +\;\left.g\!\left(\frac{m_{q'}^2}{M_{\varphi_i^0}^2}\right)\left(2{\rm Re}y_{q'}^{\varphi^0_i}{\rm Im}y_{q}^{\varphi^0_i}\right)\right\}\,,
\end{eqnarray}
where $q_f$ denotes the charge of the fermion, \ie{} $q_f=-1,-1/3,+2/3$ for $f=e,d,u$ respectively, $N_C^{f'}=3,1$ for quarks and leptons respectively, and the parametric integrals $f,g$ are given by \cite{Barr:1990vd}
\begin{eqnarray}
f(z) &\equiv& \frac{1}{2}z\int_0^1dx\frac{1-2x(1-x)}{x(1-x)-z}\log\frac{x(1-x)}{z}\,\quad{\rm and}\nonumber\\
g(z)&\equiv& \frac{1}{2}z\int_0^1dx\frac{1}{x(1-x)-z}\log\frac{x(1-x)}{z}\,.
\end{eqnarray}
These integrals are of order one for a top in the loop and scale (only) approximately linearly with the fermion mass. We include therefore apart from the top contribution also the ones from the beauty quark and the tau in the sums over $f',q'$ above.

The contribution with a charged gauge boson in the loop can be translated into our model parameters via
\begin{equation}
\sin^2\!\beta\, Z_{0n} = y_f^{\varphi^0_i}\mathcal{R}_{i1}\,,
\end{equation}
implying
\begin{equation}\label{eq::eEDMBZ0}
\frac{d_{f}^\gamma(\mu_{tH};\varphi^0,W)_{BZ}}{2} = 2q_f\,m_f\frac{\sqrt{2}G_F\alpha}{(4\pi)^3}\sum_{i}\left[3f\!\left(\frac{M_W^2}{M_{\varphi_i^0}^2}\right)+5g\!\left(\frac{M_W^2}{M_{\varphi_i^0}^2}\right)\right]{\rm Im}\!\left(y_f^{\varphi^0_i}\mathcal{R}_{i1}\right)\,.
\end{equation}
Note that again the sum of contributions cancels for degenerate Higgs masses as well as in the decoupling limit. The main difference to the contributions with quark loops is that it contains no part quadratic in the parameters $\varsigma_i$. The order of magnitude for the single contributions is that of the top-loop one, as the suppression due to the smaller mass is compensated by the larger charge. 

Note that the diagrams discussed here are only a subset of the contributing ones, see the references given above. 
However, none of the additional contributions has been found to dominate over the ones discussed here. As in \cite{Barr:1990vd} we assume that they do not exhibit strong cancellations with the ones included in our calculation.\footnote{Note, however, that \eg{} in \cite{Gunion:1990ce,Shu:2013uua} cancellations between different contributions in some part of parameter space have been observed.} Furthermore we observe that in general the CEDM contribution dominates over the EDM one. Therefore our calculation is still expected to give reasonable upper limits on the CP-violating parameters in the Yukawa sector, to the extent discussed further in Secs.~\ref{sec::neutralHiggs} and~\ref{sec::neutralHiggspheno}.

\subsubsection{Charged Higgs contribution \label{sec::BZcharged}}
As mentioned before, for CP-violating charged Higgs couplings there exist a number of corresponding diagrams not calculated in \cite{Barr:1990vd}. For the electron EDM, the contribution with CP violation stemming from the Yukawa couplings has been calculated in \cite{BowserChao:1997bb}. In general, the translation to quarks is non-trivial, as the authors give the result only for $m_\nu(=m_b)=0$, while for the quark EDM \eg{} contributions with relative weight $m_b\varsigma_d/(m_t \varsigma_u)$ could exist. However, our analysis yields that all additional contributions are either of the order $m_b^2\varsigma_d/(m_t^2 \varsigma_u),\,m_b^2\varsigma_d/(m_t M_H \varsigma_u),$ or CP-conserving, implying that the translation \emph{is} in this case possible without evaluating new diagrams. We therefore start from the result from \cite{BowserChao:1997bb},\footnote{The factor $|V_{tb}|^2\sim1$ has been omitted in that reference.} identifying $c_t=\varsigma_u$ and $c_e=-\varsigma_l$,
\begin{eqnarray}\label{eq::eEDMBZc}
\frac{d_{e}^\gamma(\mu_{tH};\varphi^\pm)_{\rm BZ}}{2} &=& -\,m_e\frac{3 g^2}{64\pi^2}\frac{g^2}{32\pi^2M_W^2}|V_{tb}|^2{\rm Im}(\varsigma_u^*\varsigma_l)(q_tF_t+q_bF_b)\nonumber\\
&=& -\,m_e\frac{12G_F^2M_W^2}{(4\pi)^4}|V_{tb}|^2{\rm Im}(\varsigma_u^*\varsigma_l)(q_tF_t+q_bF_b)
\end{eqnarray}
and use\footnote{Note that we correct here the sign for the second term in $T_t(z)$ as compared to \cite{BowserChao:1997bb}.}
\begin{eqnarray}
F_q &=& \frac{T_q(z_H)-T_q(z_W)}{z_H-z_W}\,,\quad\mbox{with}\quad z_x:=M_x^2/m_t^2\,,\\
T_t(z) &=& \frac{1-3z}{z^2}
\frac{\pi^2}{6}+\left(\frac{1}{z}-\frac{5}{2}\right)\log z-\frac{1}{z}-\left(2-\frac{1}{z}\right)\left(1-\frac{1}{z}\right){\rm Li}_2(1-z)\quad {\rm and}\nonumber\\
T_b(z) &=&\frac{2z-1}{z^2}\frac{\pi^2}{6}+\left(\frac{3}{2}-\frac{1}{z}\right)\log z+\frac{1}{z}-\frac{1}{z}\left(2-\frac{1}{z}\right){\rm Li}_2(1-z)\,.\nonumber
\end{eqnarray}
Note that the functions $F_q$ are of course finite for $M_H\to M_W$ ($F_t|_{M_H=M_W}\sim2$ and $F_b|_{M_H=M_W}\sim 1$). Furthermore, $(q_t F_t+q_b F_b)\in[0,1]$ and $\lim_{M_{H^\pm}\to\infty}F_q=0$ hold.
The generalization to the down quark reads as follows:
\begin{equation}\label{eq::BZc}
\frac{d^\gamma_{d}(\mu_{tH};\varphi^\pm)_{\rm BZ}}{2} = -\,m_d\frac{12G_F^2M_W^2}{(4\pi)^4}|V_{tb}|^2|V_{ud}|^2{\rm Im}(\varsigma_u^*\varsigma_d)(q_tF_t+q_bF_b)\,,
\end{equation}
while the up quark contribution is negligible. 

%

\section{Phenomenological Analysis\label{sec::phenomenology}}

In this section, the phenomenological analysis of the constraints discussed in the previous one is performed. Since the parametrizations in Eqs.~\eqref{eq::lagrangiancharged} and~\eqref{eq::lagrangianneutral} are general, the corresponding limits hold for any model; when the scaling differs largely from the \athdmws, however, other constraints might be stronger than the ones discussed here. In the general case, the limits concern certain matrix elements of the $\varsigma_i$, which we will indicate appropriately. The constraints given correspond to the quoted experimental limits in combination  with   extreme values of the allowed ranges for the theoretical parameters.

We start by discussing charged Higgs exchange, as in that case the interpretation of the results is straightforward. Specifically for the \athdm we can relate the results directly to those obtained in previous analyses from flavour-changing observables \cite{Jung:2010ik,Jung:2010ab}.  
Discussing the different contributions separately implies assuming that no severe cancellations occur between them, which should be kept in mind for the following discussion.
For each class, we will show in this section the most stringent constraints only; the remaining constraints are commented upon in the text.

\subsection{Charged Higgs contributions}
The contributions from charged-Higgs exchange are relevant for the neutron and electron EDMs, only; in diamagnetic systems, they are usually negligible, since they contribute neither to CEDMs nor electron-nucleon interactions. They vanish whenever the relevant factors $\varsigma_{u,d,l}$ lack a phase difference, similarly to the SM contributions.

\begin{figure}[hbt]
\centering{
\includegraphics[width=0.35\textwidth]{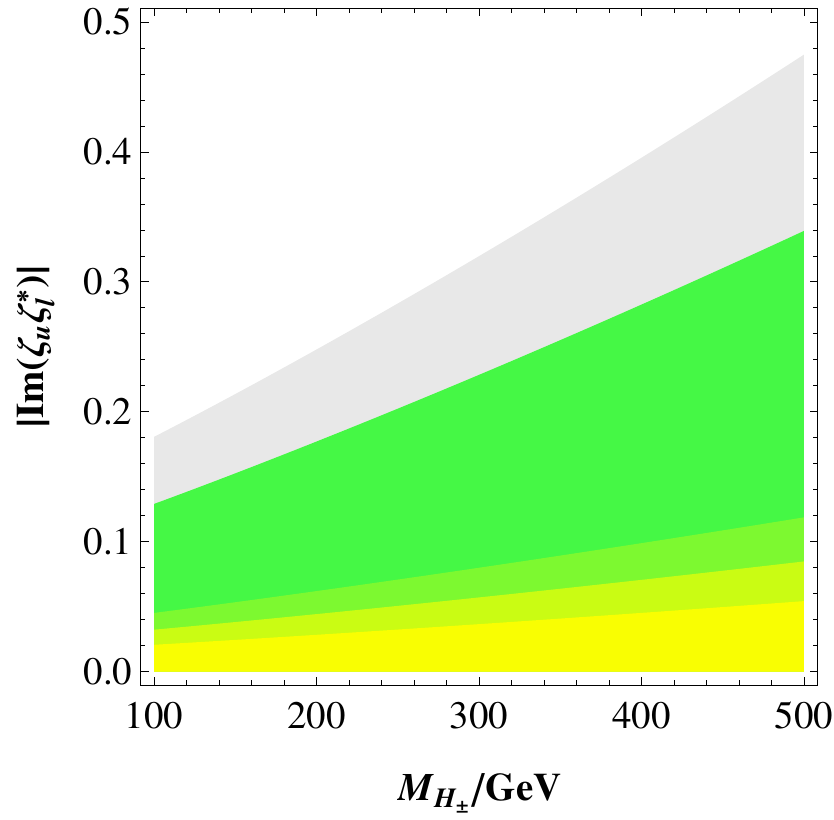}\qquad\qquad\qquad\includegraphics[width=0.35\textwidth]{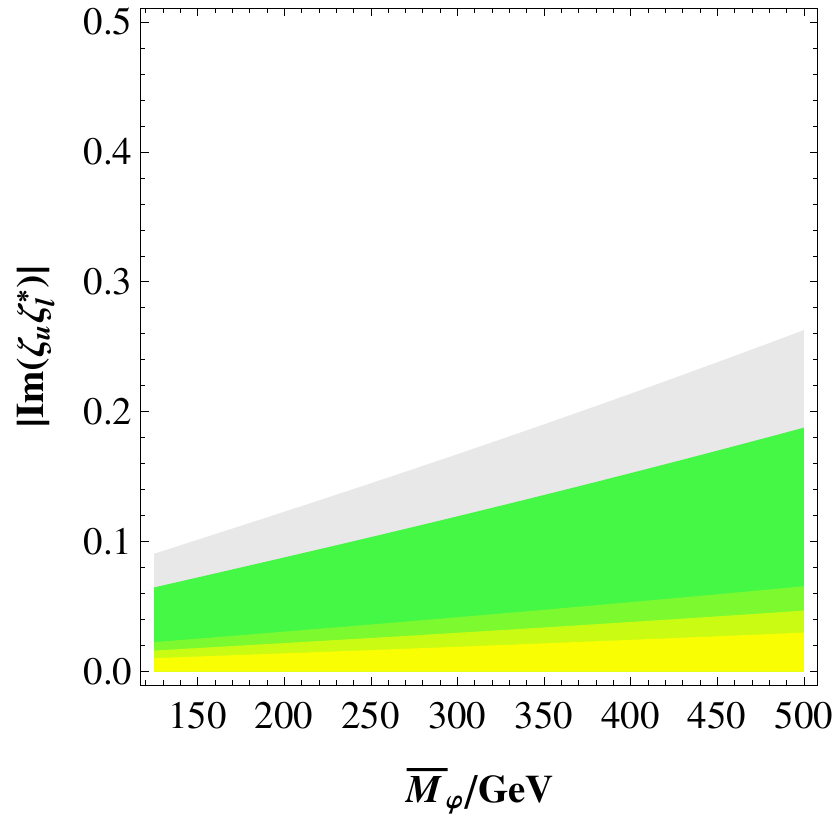}
}
\caption{The constraints from the electron EDM ($95\%$~CL) on charged Higgs exchange (left) in the ${\rm Im}(\varsigma_u^*\varsigma_l)-M_{H^\pm}$ plane and neutral Higgs exchange (right) in the ${\rm Im}(\varsigma_u^*\varsigma_l)-\overline M_{\varphi}$ plane. The grey area corresponds to the old result for $|d_e|$, the dark green one to the very conservative new fit, cf. Eq.~\eqref{eq::eEDM}. The remaining three areas correspond to $|d_e|$ obtained by making an assumption on fine-tuning ($n=1,2,3$), cf. Sec.~\ref{sec::atomEDMs}.\label{fig::eEDMzetaul}}
\end{figure}

We start by analyzing the constraint from the electron EDM as obtained in Sec.~\ref{sec::atomEDMs}. The charged Higgs contributes via Barr-Zee diagrams, cf. Eq.~(\ref{eq::eEDMBZc}); the resulting constraint is shown in Fig.~\ref{fig::eEDMzetaul} on the left, and implies $|{\rm Im}(\varsigma_u\varsigma_l^*)|\lesssim0.02-0.34$ ($\varsigma_{u,33}\varsigma_{l,11}^*$), depending on the charged scalar mass and the choice for $|d_e|$ in Table~\ref{tab::delimitnew}, together with $|{\rm Im}(\varsigma_u\varsigma_l^*)|/M_{H^\pm}^2\leq 10^{-5}~{\rm GeV}^{-2}$, to be compared with $|\varsigma_u\varsigma_l^*|/M_{H^\pm}^2\leq 10^{-2}~{\rm GeV}^{-2}$ obtained in \cite{Jung:2010ik}. This demonstrates already the strength of EDMs in constraining CP-violating parameter combinations.

The main contribution to the neutron EDM stems from the Weinberg operator, especially since there are no sizable contributions to the CEDM involving the charged Higgs. 
For the considered range of charged Higgs masses, the relative contribution from the corresponding Barr-Zee diagrams, cf. Eq.~\eqref{eq::BZc}, is about $15\%$ of the one from the Weinberg operator.

\begin{figure}[hbt]
\centering{
\includegraphics[width=0.35\textwidth]{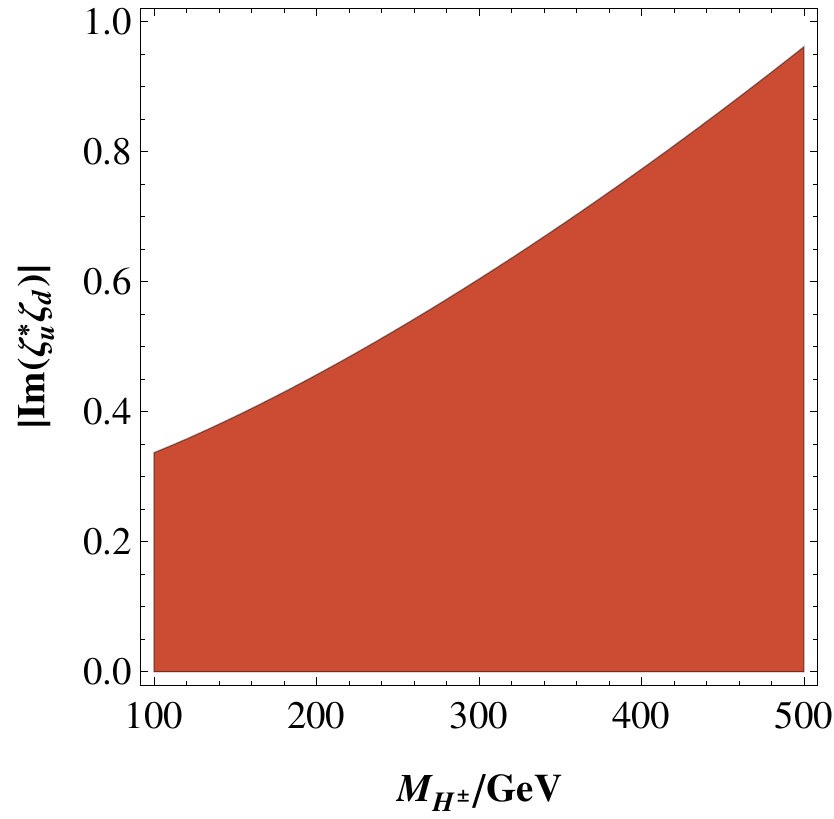}\qquad\qquad\qquad\includegraphics[width=0.35\textwidth]{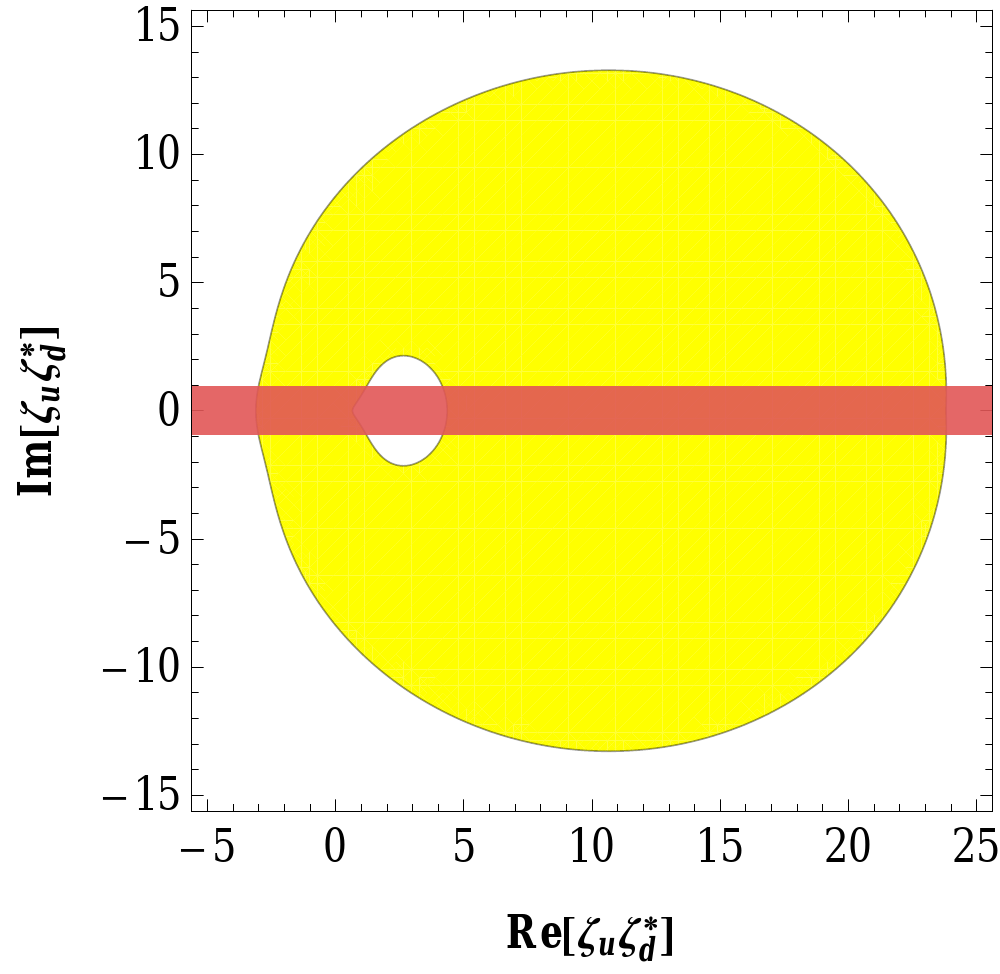}
}
\caption{The constraint from the neutron EDM ($95\%$~CL) in the ${\rm Im}(\varsigma_u^*\varsigma_d)-M_{H^\pm}$ plane (left) and together with the constraint from $BR(b\to s\gamma)$  in the complex $\varsigma_u\varsigma_d^*$ plane (right), allowing for $80~{\rm GeV}\leq M_H\leq 500~{\rm GeV}$.\label{fig::nEDMWzetaud}}
\end{figure}

Using Eqs.~\eqref{eq::runningW} and~\eqref{eq::C2}, we plot the resulting constraint in the ${\rm Im}(\varsigma_u^*\varsigma_d)-M_{H^\pm}$ plane ($\varsigma_{u,33}^*\varsigma_{d,33}$) in Fig.~\ref{fig::nEDMWzetaud}. For a charged-Higgs mass of $\sim500$~GeV, ${\rm Im}(\varsigma_u^*\varsigma_d)\lesssim1$ remains allowed, which is strengthened to $\sim0.3$ for light masses. We emphasize that therefore no fine-tuning is necessary to avoid this bound; however, the next-generation experiments will put this scenario to a non-trivial test, \ie{} we would generally expect contributions within the projected sensitivity.

To illustrate the impact of this bound in the \athdmws, we show on the right-hand side the comparison to the one arising from the branching ratio for $b\to s\gamma$ \cite{Jung:2010ab} 
in the complex $\varsigma_u\varsigma_d^*$ plane, an observable known for its high sensitivity to a second Higgs doublet. While an imaginary part of $\mathcal{O}(1)$ is still possible, it follows from the discussion in \cite{Jung:2010ab} that large effects in other observables like $A_{\mathrm{CP}}(b\to s\gamma)$ are excluded by this constraint.

\subsection{Neutral Higgs contributions\label{sec::neutralHiggspheno}}
As discussed in Sec.~\ref{sec::neutralHiggs}, the contributions from neutral scalars are more involved phenomenologically. Specifically, cancellations are likely to play an important role, cf. Eq.~\eqref{eq::ycancellation}. Since these cancellations take place for both limiting cases, universal Higgs masses and decoupling, and furthermore the mixing into the lightest mass eigenstate is rather small, see the discussion in Sec.~\ref{sec::neutralHiggs} and in \cite{Celis:2013rcs,Pich:2013vta}, we use the right-hand side of Eq.~\eqref{eq::ycancellation} as an approximation of the appearing sums. However, since the two limits imply different patterns for the single contributions, we evaluate the mass-dependent functions for a varying effective mass $\overline M_\varphi$, allowing the corresponding coefficient to take any value between the two limiting ones. That is, we have 
\begin{equation}\label{eq:prescription}
\sum_i f(M_{\varphi_i^0}){\rm Re}\left(y_f^{\varphi^0_i}\right){\rm Im}\left(y_{f'}^{\varphi^0_i}\right)\quad\to\quad \xi_{f'}\;f(\overline M_\varphi){\rm Im}\left[(\varsigma_{F(f)}^*)_{ff}(\varsigma_{F(f')})_{f'f'}\right]\,.
\end{equation}
The constraints shown can be translated back into the corresponding parameter combinations whenever a specific model with known Higgs masses is discussed.

For neutral Higgs exchanges between different families, the resulting constraints allow for a comparison with the charged Higgs contributions, albeit with some caution. Note that the two contributing terms, ${\rm Re}(y_f^{\varphi_i^0}){\rm Im}(y_{f'}^{\varphi_i^0})$ and ${\rm Re}(y_{f'}^{\varphi_i^0}){\rm Im}(y_{f}^{\varphi_i^0})$ both translate in the \athdm to ${\rm Im}(\varsigma_f\varsigma_{f'}^*)$, but for $f,f'=d,l$ with opposite signs, implying further cancellations, since the coefficient functions given in the previous section have identical signs. If one of the involved fermions is an up-type quark, the two contributions instead strengthen the bound.

For the cases in which the right-hand side vanishes, we provide the value of the contribution from the lightest scalar as a reference, which is not to be understood as a conservative limit of any kind, but as the strongest obtainable limit for the corresponding couplings in a specific model.

For neutral scalars, we do not include the contributions from the Weinberg operator, for two reasons: first they are subject to the cancellations discussed above (even in the most general case), second they are slightly smaller than the contributions from Barr-Zee diagrams with the same coefficients (for universal $\varsigma_{u,d}$), which enter now via the chromomagnetic moments.

We start again with the constraints from the electron EDM. 
In Fig.~\ref{fig::eEDMzetaul} on the right, the constraint for $|{\rm Im}(\varsigma_l\varsigma_u^*)|$ ($\varsigma_{l,11}\varsigma_{u,33}^*$) is displayed, plotted against $\overline M_{\varphi}$. This parameter combination is now bound to be $\lesssim 0.01-0.2$, depending on $M_{H^\pm}$ and the choice for $|d_e|$; this is about a factor $2$ stronger than the charged-Higgs constraint for $M_{H^\pm}\sim\overline M_\varphi$, as can also be deduced directly from Eqs.~\eqref{eq::eEDMBZ0} and \eqref{eq::eEDMBZc}. While this again does not yet call for severe fine-tuning of the parameters at the moment, the bounds are strong already, especially when accepting the bounds from ThO with restricted fine-tuning. Clearly, the coming experiments, see Table~\ref{tab::ExpEDM}, will explore a region of parameter space in which we would generally expect a signal. 
The contribution with a tau lepton in the loop, proportional to ${\rm Im}(y_l^2)$ (${\rm Re}(y_{l,11}){\rm Im}(y_{l,33})$ and ${\rm Re}(y_{l,33}){\rm Im}(y_{l,11})$), is subject to strong cancellations in the \athdmws; there is therefore no conservative limit. The contribution at the lightest Higgs mass yields $|{\rm Im}(y_l^2)|/2\leq 2-15$, depending on $|d_e|$.
For a beauty quark in the loop, the constraint is weaker than the one obtained from the bound on the electron-nucleon coupling $\tilde C_S$; it is therefore omitted. It is noteworthy, however, that the single contributions are smaller than the ones with the tau in the loop, despite the larger mass of the beauty quark, due to the smaller charge and the occurring cancellation.
Finally, the gauge boson loops give potentially large contributions, however again subject to strong cancellations. Furthermore,
since the admixture of the lightest mass eigenstate with the second doublet is small, this contribution gets further suppressed. Having this in mind, however, the contribution from the lightest neutral scalar yields $|\mathcal{R}_{11}{\rm Im}(y_l^{\varphi_1^0})|\leq0.01-0.07$, again depending on the value for $|d_e|$.

\begin{figure}[hbt!]
\centering{
\includegraphics[width=0.35\textwidth]{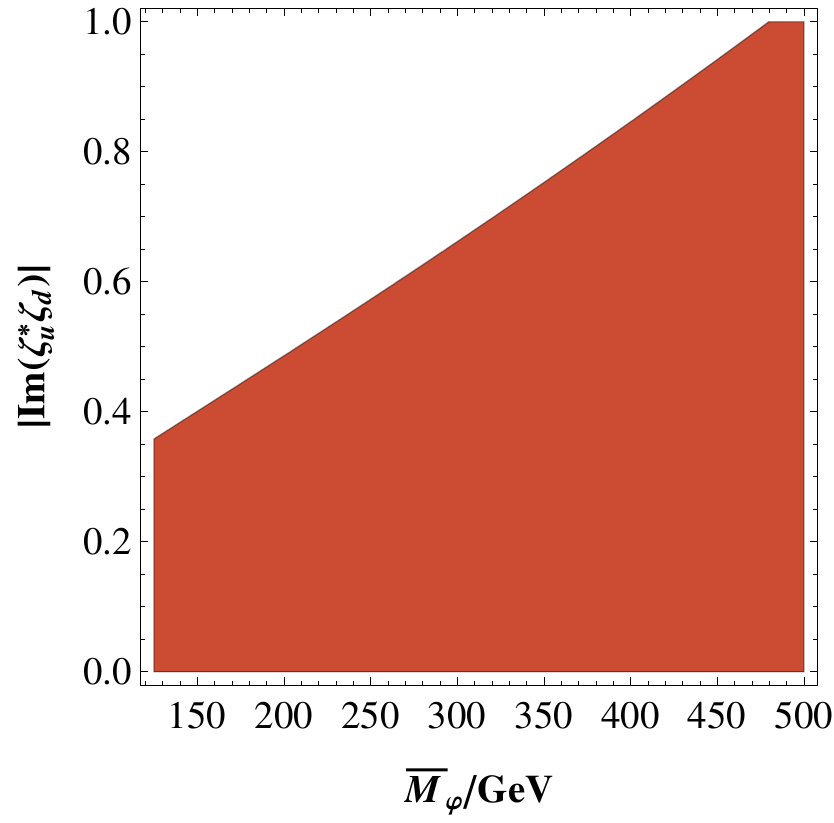}}
\caption{Constraint from the neutron EDM ($95\%$~CL) in the $|{\rm Im}(\varsigma_d\varsigma_u^*)|-\overline{M}_\varphi$ plane. \label{fig::nEDMBZyud}}
\end{figure}

The main constraint from the neutron EDM is again for the combination $\varsigma_u^*\varsigma_d$ ($\varsigma_{u,33}\varsigma_{d,11}$), since it is enhanced by the top mass and involves different families. The resulting constraint, shown in Fig.~\ref{fig::nEDMBZyud}, is similar to the one from the charged Higgs exchange via the Weinberg operator; given this situation, the treatment for the hadronic matrix element is decisive for their relative strength and possible cancellations.
The other constraints are either again subject to strong cancellations ($|{\rm Im}(y_u^2)|/2\leq 1.4$, $|{\rm Im}(y_d^2)|/2\leq 26$ and $|\mathcal{R}_{11}{\rm Im}(y_d^{\phi_1^0})|\leq 3.6$ for the lightest scalar) or not constraining due to the small masses involved.

\begin{figure}[hbt!]
\centering{
\includegraphics[width=0.35\textwidth]{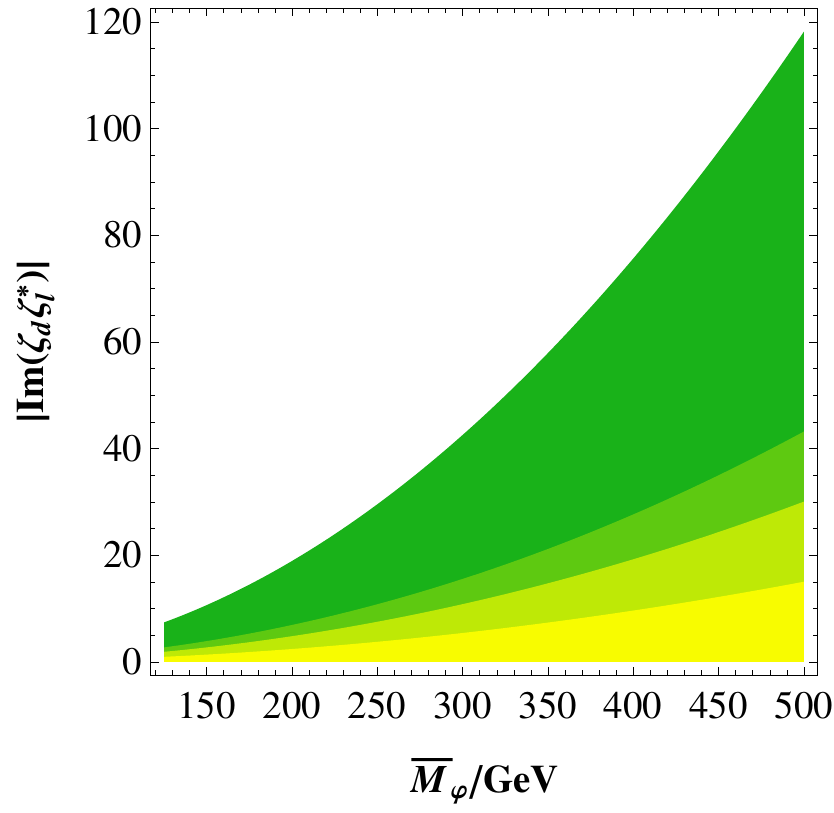}\qquad\qquad\qquad\includegraphics[width=0.35\textwidth]{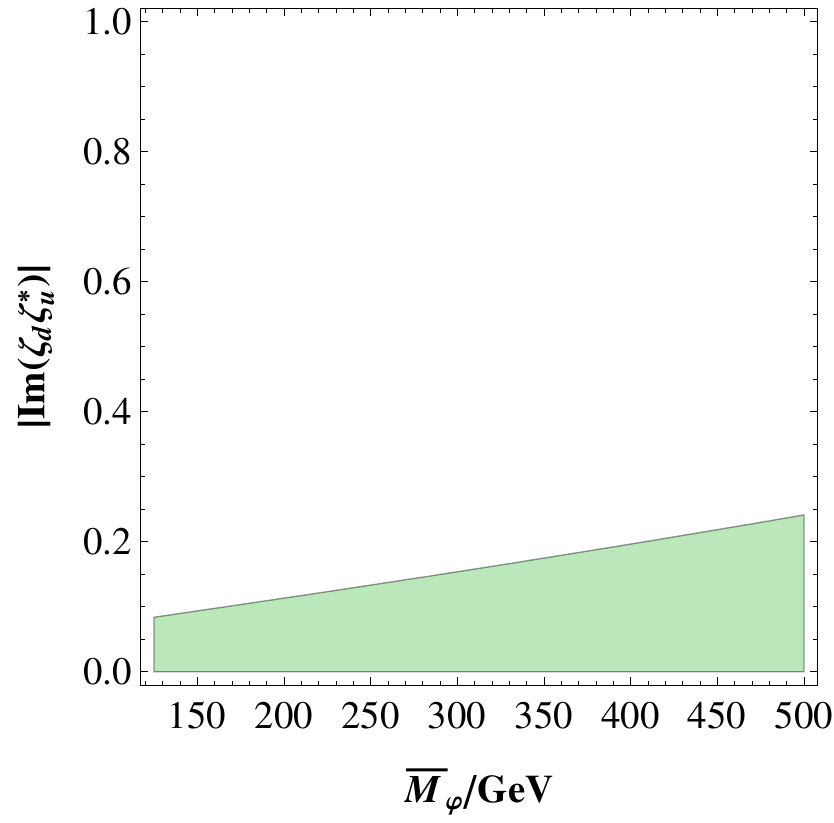}
}
\caption{Actual constraint from the mercury EDM ($95\%$~CL) in the $|{\rm Im}(\varsigma_d\varsigma_l^*)|-\overline{M}_\varphi$ plane (left) for the old bound on $|\tilde C_S|$ as well as the ones from ThO, and the potential constraint from the same system in the $|{\rm Im}(\varsigma_d\varsigma_u^*)|-M_\varphi$ plane (right), see text.\label{fig::HgEDMydl}}
\end{figure}

The final constraints we consider stem from the mercury EDM. As discussed above, relating this observable to fundamental parameters is complicated by large theory uncertainties. However, \eg{}  the electron-nucleon couplings are not as strongly affected by these uncertainties, providing a more reliable bound. Furthermore, it is a conservative one: this contribution is not expected to be dominating this observable; for that reason, assuming this contribution to saturate the experimental limit is conservative. This fact has been used in \cite{Jung:2013mg} to obtain the limit $\tilde C_S\leq 7\times10^{-8}$ (mainly) from the mercury measurement, thereby allowing for a model-independent limit on the electron EDM. Here, as we are expressing both $\tilde C_S$ and $\tilde C_P$ by coefficients of four-fermion operators, we make this assumption for their combination appearing in Eq.~\eqref{eq::HgEDM}; additionally, we show the bounds from ThO with the fine-tuning assumption. The resulting constraints are  shown in Fig.~\ref{fig::HgEDMydl} on the left. They do not appear very strong numerically, but constrain a parameter combination which was allowed to be very large before and are therefore relevant. Note that the contribution from $\tilde C_P$ weakens slightly the constraint compared to using $\tilde C_S\leq 7\times 10^{-8}$, but not severely. 

Further contributions enter via the (Barr-Zee-)CEDM contributions to the Schiff moment, yielding potentially strong bounds; we illustrate their potential impact by using simply central values for the hadronic parameters in the equations above to obtain a constraint. 
An exemplary result, corresponding to a  more reliable theoretical situation, is shown in Fig.~\ref{fig::HgEDMydl} on the right. We note that for the assumed situation, it would be the strongest limit on $|{\rm Im}(\varsigma_d\varsigma_u^*)|$ available.
Furthermore, the physical mechanisms are different for the various systems. Specifically, for mercury the charged Higgs plays a minor role, so a possible cancellation for the neutron between these two contributions cannot take place here. Theoretical progress for this observable would therefore be very valuable.

\section{Conclusions\label{sec::conclusions}}
EDMs are very sensitive probes of NP models incorporating additional sources of
CP violation. In particular, they strongly constrain possible new flavour-blind phases, as those present in generic 2HDMs without tree-level FCNCs.
We have critically analyzed the present experimental limits on EDMs of elementary particles and composite systems (nucleons, nuclei, atoms and molecules), and have derived the resulting phenomenological constraints on the 2HDM parameters.

To be specific, our final results are written in the context of the A2HDM, where the alignment in flavour space of the two Yukawa matrices coupling to a given right-handed fermion guarantees the absence of tree-level FCNCs, while allowing for flavour-blind Yukawa phases. This theoretical framework includes (and generalizes) all particular (CP-conserving) types of 2HDMs based on discrete $\mathcal{Z}_2$ symmetries, usually adopted in the literature. Nevertheless, our findings can be directly applied to even more general Yukawa structures with simple notational changes.

The symmetries of the A2HDM protect in a very efficient way the flavour-blind phases from undesirable phenomenological consequences. Although the present experimental limits impose indeed strong bounds on the CP-violating parameter combinations, $\cO(1)$ contributions remain allowed. However, large enhancements in other CP-violating observables are already strongly restricted by the present EDM bounds.

A strong caveat to keep in mind is the strong sensitivity of the EDM predictions to the UV completion of the low-energy 2HDM. Since the A2HDM flavour symmetries strongly suppress any possible tree-level or one-loop contribution, the predicted EDMs originate from two-loop diagrams. Therefore, these theoretical results could easily be changed by NP contributions beyond the 2HDM, as happens for instance in supersymmetry, and unexpected cancellations could also take place. The EDM constraints should then be interpreted with a lot of care.

Within the A2HDM, the dominant mechanisms generating non-zero EDMs are charged and neutral scalar exchanges through two-loop diagrams of the Weinberg and Barr-Zee type. While the charged Higgs contributions can be determined unambiguously, the mixing among the three neutral scalars makes their effect much more subtle. We have shown that the neutral scalar contribution for a given fermion species vanishes exactly in two opposite limits: universal Higgs masses and decoupling. The null result is due to the orthogonality of the scalar mixing matrix, which generates exact cancellations among the contributions from the three neutral scalars. This fact has been sometimes overlooked in the literature, leading to claims of non-vanishing contributions in the decoupling limit which are not correct in this context. In particular, simply taking the contribution from the lightest Higgs is not necessarily a good approximation.
In order to obtain a phenomenological estimate of the neutral scalar effect we have taken an average scalar mass to evaluate any mass-dependent function and followed the prescription indicated in Eq.~(\ref{eq:prescription}); we have only provided as a reference the value of the contribution from the lightest scalar
in those cases where the right-hand side of Eq.~(\ref{eq:prescription}) vanishes.

Our final phenomenological results are shown in Figs. 2 to 5. In spite of all previous comments of caution, these plots indicate that interesting signals could be expected within the projected sensitivity of the next-generation of EDM experiments. Experimental progress in this field could then bring a break-through in the search for NP phenomena.

\section*{Acknowledgements}
The authors acknowledge Paula Tuz\'on for collaboration during early stages of the project. M.J. would like to thank Vladimir Braun, Jorge Martin Camalich, Bhanu Das, Vladimir Dzuba, Jon Engel, Celal Harabati, Dilip Kumar Singh and Adam Ritz for helpful discussions. This work has been supported in part by the Spanish Government and EU funds for regional development [grants FPA2007-60323, FPA2011-23778 and CSD2007-00042 (Consolider Project CPAN)], and the Generalitat Valenciana [PrometeoII/2013/007]. 
The work of M.J. is funded by the german Federal Ministry of Education and Research (BMBF).

\bibliography{EDM_v2}

\begin{thebibliography}{100}

\bibitem{Sakharov:1967dj}
A.D. Sakharov.
\newblock {Violation of CP Invariance, c Asymmetry, and Baryon Asymmetry of the
  Universe}.
\newblock {\em Pisma Zh.Eksp.Teor.Fiz.}, 5:32--35, 1967.

\bibitem{Pospelov:2005pr}
Maxim Pospelov and Adam Ritz.
\newblock {Electric dipole moments as probes of new physics}.
\newblock {\em Annals Phys.}, 318:119--169, 2005.

\bibitem{Mannel:2012qk}
Thomas Mannel and Nikolai Uraltsev.
\newblock {Loop-Less Electric Dipole Moment of the Nucleon in the Standard
  Model}.
\newblock {\em Phys.Rev.}, D85:096002, 2012.

\bibitem{Peccei:1977hh}
R.D. Peccei and Helen~R. Quinn.
\newblock {CP Conservation in the Presence of Instantons}.
\newblock {\em Phys.Rev.Lett.}, 38:1440--1443, 1977.

\bibitem{Weinberg:1977ma}
Steven Weinberg.
\newblock {A New Light Boson?}
\newblock {\em Phys. Rev. Lett.}, 40:223--226, 1978.

\bibitem{Wilczek:1977pj}
Frank Wilczek.
\newblock {Problem of Strong P and T Invariance in the Presence of Instantons}.
\newblock {\em Phys. Rev. Lett.}, 40:279--282, 1978.

\bibitem{Kobayashi}
M.~Kobayashi and T.~Maskawa.
\newblock {CP violation in the renormalizable theory of weak interaction}.
\newblock {\em Prog. Theor. Phys.}, 49:652, 1973.

\bibitem{Weinberg:1989dx}
Steven Weinberg.
\newblock {Larger Higgs Exchange Terms in the Neutron Electric Dipole Moment}.
\newblock {\em Phys. Rev. Lett.}, 63:2333, 1989.

\bibitem{Glashow:1976nt}
Sheldon~L. Glashow and Steven Weinberg.
\newblock {Natural Conservation Laws for Neutral Currents}.
\newblock {\em Phys.Rev.}, D15:1958, 1977.

\bibitem{Paschos:1976ay}
E.A. Paschos.
\newblock {Diagonal Neutral Currents}.
\newblock {\em Phys.Rev.}, D15:1966, 1977.

\bibitem{Cheng:1987rs}
T.~P. Cheng and Marc Sher.
\newblock {Mass Matrix Ansatz and Flavor Nonconservation in Models with
  Multiple Higgs Doublets}.
\newblock {\em Phys. Rev.}, D35:3484, 1987.

\bibitem{Branco:1996bq}
G.C. Branco, W.~Grimus, and L.~Lavoura.
\newblock {Relating the scalar flavor changing neutral couplings to the CKM
  matrix}.
\newblock {\em Phys.Lett.}, B380:119--126, 1996.

\bibitem{Atwood:1996vj}
David Atwood, Laura Reina, and Amarjit Soni.
\newblock {Phenomenology of two Higgs doublet models with flavor changing
  neutral currents}.
\newblock {\em Phys. Rev.}, D55:3156--3176, 1997.

\bibitem{DiazCruz:2009ek}
J.~L. Diaz-Cruz, J.~Hernandez-Sanchez, S.~Moretti, R.~Noriega-Papaqui, and
  A.~Rosado.
\newblock {Yukawa Textures and Charged Higgs Boson Phenomenology in the
  2HDM-III}.
\newblock {\em Phys. Rev.}, D79:095025, 2009.

\bibitem{Pich:2009sp}
Antonio Pich and Paula Tuzon.
\newblock {Yukawa Alignment in the Two-Higgs-Doublet Model}.
\newblock {\em Phys. Rev.}, D80:091702, 2009.

\bibitem{Botella:2009pq}
F.J. Botella, G.C. Branco, and M.N. Rebelo.
\newblock {Minimal Flavour Violation and Multi-Higgs Models}.
\newblock {\em Phys.Lett.}, B687:194--200, 2010.

\bibitem{Jung:2010ik}
Martin Jung, Antonio Pich, and Paula Tuzon.
\newblock {Charged-Higgs phenomenology in the Aligned two-Higgs- doublet
  model}.
\newblock {\em JHEP}, 11:003, 2010.

\bibitem{Buras:2010mh}
Andrzej~J. Buras, Maria~Valentina Carlucci, Stefania Gori, and Gino Isidori.
\newblock {Higgs-mediated FCNCs: Natural Flavour Conservation vs. Minimal
  Flavour Violation}.
\newblock {\em JHEP}, 1010:009, 2010.

\bibitem{Varzielas:2011jr}
Ivo de~Medeiros~Varzielas.
\newblock {Family symmetries and alignment in multi-Higgs doublet models}.
\newblock {\em Phys.Lett.}, B701:597--600, 2011.

\bibitem{Crivellin:2013wna}
Andreas Crivellin, Christoph Greub, and Ahmet Kokulu.
\newblock Flavor-phenomenology of two-higgs-doublet models with generic yukawa
  structure.
\newblock {\em Phys. Rev. D}, 87:094031, 2013.

\bibitem{Shu:2013uua}
Jing Shu and Yue Zhang.
\newblock {Impact of a CP Violating Higgs: from LHC to Baryogenesis}.
\newblock {\em Phys.Rev.Lett.}, 111:091801, 2013.

\bibitem{Branco:2011iw}
G.C. Branco, P.M. Ferreira, L.~Lavoura, M.N. Rebelo, Marc Sher, et~al.
\newblock {Theory and phenomenology of two-Higgs-doublet models}.
\newblock {\em Phys.Rept.}, 516:1--102, 2012.

\bibitem{Jung:2013mg}
Martin Jung.
\newblock {A robust limit for the electric dipole moment of the electron}.
\newblock {\em JHEP}, 1305:168, 2013.

\bibitem{Jung:2010ab}
Martin Jung, Antonio Pich, and Paula Tuzon.
\newblock {The B $\to$ Xs gamma Rate and CP Asymmetry within the Aligned
  Two-Higgs-Doublet Model}.
\newblock {\em Phys. Rev.}, D83:074011, 2011.

\bibitem{Jung:2012vu}
Martin Jung, Xin-Qiang Li, and Antonio Pich.
\newblock {Exclusive radiative B-meson decays within the aligned
  two-Higgs-doublet model}.
\newblock {\em JHEP}, 1210:063, 2012.

\bibitem{Celis:2012dk}
Alejandro Celis, Martin Jung, Xin-Qiang Li, and Antonio Pich.
\newblock {Sensitivity to charged scalars in $B\to D^{(*)}\tau\nu_\tau$ and
  $B\to\tau\nu_\tau$ decays}.
\newblock {\em JHEP}, 1301:054, 2013.

\bibitem{Celis:2013rcs}
Alejandro Celis, Victor Ilisie, and Antonio Pich.
\newblock {LHC constraints on two-Higgs doublet models}.
\newblock {\em JHEP}, 1307:053, 2013.

\bibitem{Duarte:2013zfa}
Lucia Duarte, Gabriel~A. González-Sprinberg, and Jordi Vidal.
\newblock {Top quark anomalous tensor couplings in the two-Higgs-doublet
  models}.
\newblock {\em JHEP}, 1311:114, 2013.

\bibitem{Ginges:2003qt}
J.~S.~M. Ginges and V.~V. Flambaum.
\newblock {Violations of fundamental symmetries in atoms and tests of
  unification theories of elementary particles}.
\newblock {\em Phys. Rept.}, 397:63--154, 2004.

\bibitem{Raidal:2008jk}
M.~Raidal et~al.
\newblock {Flavour physics of leptons and dipole moments}.
\newblock {\em Eur. Phys. J.}, C57:13--182, 2008.

\bibitem{Fukuyama:2012np}
Takeshi Fukuyama.
\newblock {Searching for New Physics beyond the Standard Model in Electric
  Dipole Moment}.
\newblock {\em Int.J.Mod.Phys.}, A27:1230015, 2012.

\bibitem{Engel:2013lsa}
Jonathan Engel, Michael~J. Ramsey-Musolf, and U.~van Kolck.
\newblock {Electric Dipole Moments of Nucleons, Nuclei, and Atoms: The Standard
  Model and Beyond}.
\newblock {\em Prog.Part.Nucl.Phys.}, 71:21--74, 2013.

\bibitem{Ellis:2008zy}
John~R. Ellis, Jae~Sik Lee, and Apostolos Pilaftsis.
\newblock {Electric Dipole Moments in the MSSM Reloaded}.
\newblock {\em JHEP}, 10:049, 2008.

\bibitem{Ellis:2011hp}
John Ellis, Jae~Sik Lee, and Apostolos Pilaftsis.
\newblock {Maximal Electric Dipole Moments of Nuclei with Enhanced Schiff
  Moments}.
\newblock {\em JHEP}, 02:045, 2011.

\bibitem{Hisano:2008hn}
Junji Hisano, Minoru Nagai, and Paride Paradisi.
\newblock {Flavor effects on the electric dipole moments in supersymmetric
  theories: A beyond leading order analysis}.
\newblock {\em Phys. Rev.}, D80:095014, 2009.

\bibitem{Hisano:2006mj}
Junji Hisano, Minoru Nagai, and Paride Paradisi.
\newblock {New two-loop contributions to hadronic EDMs in the MSSM}.
\newblock {\em Phys. Lett.}, B642:510--517, 2006.

\bibitem{Altmannshofer:2009ne}
Wolfgang Altmannshofer, Andrzej~J. Buras, Stefania Gori, Paride Paradisi, and
  David~M. Straub.
\newblock {Anatomy and Phenomenology of FCNC and CPV Effects in SUSY Theories}.
\newblock {\em Nucl. Phys.}, B830:17--94, 2010.

\bibitem{Li:2010ax}
Yingchuan Li, Stefano Profumo, and Michael Ramsey-Musolf.
\newblock {A Comprehensive Analysis of Electric Dipole Moment Constraints on
  CP-violating Phases in the MSSM}.
\newblock {\em JHEP}, 08:062, 2010.

\bibitem{Mercolli:2009ns}
Lorenzo Mercolli and Christopher Smith.
\newblock {EDM constraints on flavored CP-violating phases}.
\newblock {\em Nucl. Phys.}, B817:1--24, 2009.

\bibitem{Ilakovac:2013wfa}
Amon Ilakovac, Apostolos Pilaftsis, and Luka Popov.
\newblock {Lepton Dipole Moments in Supersymmetric Low-Scale Seesaw Models}.
\newblock {\em Phys.Rev.}, D89:015001, 2014.

\bibitem{Dhuria:2013ida}
Mansi Dhuria and Aalok Misra.
\newblock {A Healthy Electron/Neutron EDM in D3/D7 mu-Split SUSY}.
\newblock 2013.
\newblock arXiv:1308.3233 [hep-ph].

\bibitem{Trott:2010iz}
Michael Trott and Mark~B. Wise.
\newblock {On theories of enhanced CP violation in $B_s,d$ meson mixing}.
\newblock {\em JHEP}, 11:157, 2010.

\bibitem{Buras:2010zm}
Andrzej~J. Buras, Gino Isidori, and Paride Paradisi.
\newblock {EDMs vs. CPV in $B_{s,d}$ mixing in two Higgs doublet models with
  MFV}.
\newblock {\em Phys. Lett.}, B694:402--409, 2011.

\bibitem{Batell:2010qw}
Brian Batell and Maxim Pospelov.
\newblock {Bs Mixing and Electric Dipole Moments in MFV}.
\newblock {\em Phys. Rev.}, D82:054033, 2010.

\bibitem{D'Ambrosio:2002ex}
G.~D'Ambrosio, G.~F. Giudice, G.~Isidori, and A.~Strumia.
\newblock {Minimal flavour violation: An effective field theory approach}.
\newblock {\em Nucl. Phys.}, B645:155--187, 2002.

\bibitem{Khriplovich:1997ga}
I.B. Khriplovich and S.K. Lamoreaux.
\newblock {\em {CP violation without strangeness: Electric dipole moments of
  particles, atoms, and molecules}}.
\newblock Springer-Verlag Berlin Heidelberg New York, 1997.

\bibitem{Manohar:1983md}
Aneesh Manohar and Howard Georgi.
\newblock {Chiral Quarks and the Nonrelativistic Quark Model}.
\newblock {\em Nucl. Phys.}, B234:189, 1984.

\bibitem{Bigi:1991rh}
Ikaros I.~Y. Bigi and N.~G. Uraltsev.
\newblock {Effective gluon operators and the dipole moment of the neutron}.
\newblock {\em Sov. Phys. JETP}, 73:198--210, 1991.

\bibitem{Bernard:2007zu}
Veronique Bernard.
\newblock {Chiral Perturbation Theory and Baryon Properties}.
\newblock {\em Prog.Part.Nucl.Phys.}, 60:82--160, 2008.

\bibitem{deVries:2010ah}
J.~de~Vries, R.~G.~E. Timmermans, E.~Mereghetti, and U.~van Kolck.
\newblock {The Nucleon Electric Dipole Form Factor From Dimension-Six
  Time-Reversal Violation}.
\newblock {\em Phys. Lett.}, B695:268--274, 2011.

\bibitem{Baron:2013eja}
Jacob Baron et~al.
\newblock {Order of Magnitude Smaller Limit on the Electric Dipole Moment of
  the Electron}.
\newblock {\em Science Magazine}, 343 (6168):269--272, 2014.

\bibitem{Pospelov:2000bw}
Maxim Pospelov and Adam Ritz.
\newblock {Neutron EDM from electric and chromoelectric dipole moments of
  quarks}.
\newblock {\em Phys. Rev.}, D63:073015, 2001.

\bibitem{Hisano:2012sc}
Junji Hisano, Jeong~Yong Lee, Natsumi Nagata, and Yasuhiro Shimizu.
\newblock {Reevaluation of Neutron Electric Dipole Moment with QCD Sum Rules}.
\newblock {\em Phys.Rev.}, D85:114044, 2012.

\bibitem{Aoki:2008ku}
Y.~Aoki et~al.
\newblock {Proton lifetime bounds from chirally symmetric lattice QCD}.
\newblock {\em Phys.Rev.}, D78:054505, 2008.

\bibitem{Braun:2008ur}
Vladimir~M. Braun et~al.
\newblock {Nucleon distribution amplitudes and proton decay matrix elements on
  the lattice}.
\newblock {\em Phys.Rev.}, D79:034504, 2009.

\bibitem{Gruber:2010bj}
Michael Gruber.
\newblock {The nucleon wave function at the origin}.
\newblock {\em Phys.Lett.}, B699:169--173, 2011.

\bibitem{Jin:1993nn}
Xue-min Jin, Marina Nielsen, and J.~Pasupathy.
\newblock {Nucleon sigma term from QCD sum rule}.
\newblock {\em Phys.Lett.}, B314:163--167, 1993.

\bibitem{Ritzprivcomm}
Adam Ritz, priv. comm.

\bibitem{Fuyuto:2012yf}
Kaori Fuyuto, Junji Hisano, and Natsumi Nagata.
\newblock {Neutron Electric Dipole Moment Induced by the Strangeness
  Revisited}.
\newblock {\em Phys.Rev.}, D87:054018, 2013.

\bibitem{GellMann:1968rz}
Murray Gell-Mann, R.J. Oakes, and B.~Renner.
\newblock {Behavior of current divergences under SU(3) x SU(3)}.
\newblock {\em Phys.Rev.}, 175:2195--2199, 1968.

\bibitem{Demir:2002gg}
Durmus~A. Demir, Maxim Pospelov, and Adam Ritz.
\newblock {Hadronic EDMs, the Weinberg operator, and light gluinos}.
\newblock {\em Phys. Rev.}, D67:015007, 2003.

\bibitem{Demir:2003js}
Durmus~A. Demir, Oleg Lebedev, Keith~A. Olive, Maxim Pospelov, and Adam Ritz.
\newblock {Electric dipole moments in the MSSM at large tan(beta)}.
\newblock {\em Nucl. Phys.}, B680:339--374, 2004.

\bibitem{Schiff:1963zz}
L.~I. Schiff.
\newblock {Measurability of Nuclear Electric Dipole Moments}.
\newblock {\em Phys. Rev.}, 132:2194--2200, 1963.

\bibitem{Sandars:1965xx}
P.~G.~H. Sandars.
\newblock The electric dipole moment of an atom.
\newblock {\em Physics Letters}, 14(3):194, 1965.

\bibitem{Sandars:1966xx}
P.~G.~H. Sandars.
\newblock Enhancement factor for the electric dipole moment of the valence
  electron in an alkali atom.
\newblock {\em Physics Letters}, 22(3):290--291, 1966.

\bibitem{Flambaum:1976vg}
V.V Flambaum.
\newblock {On Enhancement of the electron Electric Dipole Moment in Heavy
  Atoms}.
\newblock {\em Yad.Fiz.}, 24:383--386, 1976.

\bibitem{Dzuba:2009kn}
V.A. Dzuba, V.V. Flambaum, and S.G. Porsev.
\newblock {Calculation of P,T-odd electric dipole moments for diamagnetic atoms
  Xe-129, Yb-171, Hg-199, Rn-211, and Ra-225}.
\newblock {\em Phys.Rev.}, A80:032120, 2009.

\bibitem{PhysRevA.78.010502}
Edmund~R. Meyer and John~L. Bohn.
\newblock Prospects for an electron electric-dipole moment search in metastable
  tho and $\mathrm{Th}{\mathrm{f}}^{+}$.
\newblock {\em Phys. Rev. A}, 78:010502, 2008.

\bibitem{2013arXiv1308.0414S}
L.~V. {Skripnikov}, A.~N. {Petrov}, and A.~V. {Titov}.
\newblock {Theoretical study of ThO for the electron electric dipole moment
  search}.
\newblock {\em arXiv: 1308.0414}, 2013.

\bibitem{Fleig:2014uaa}
Timo Fleig and Malaya~K. Nayak.
\newblock {Electron Electric Dipole Moment and Hyperfine Interaction Constants
  for ThO}.
\newblock 2014.
\newblock arXiv:1401.2284 [physics.atom-ph].

\bibitem{PhysRevA.85.029901}
V.~A. Dzuba, V.~V. Flambaum, and C.~Harabati.
\newblock Relations between matrix elements of different weak interactions and
  interpretation of the parity-nonconserving and electron
  electric-dipole-moment measurements in atoms and molecules.
\newblock {\em Phys. Rev. A}, 84:052108, 2011.
\newblock Erratum ibid, {\bf 85}, 029901 (2012).

\bibitem{Regan:2002ta}
B.~C. Regan, E.~D. Commins, C.~J. Schmidt, and D.~DeMille.
\newblock {New limit on the electron electric dipole moment}.
\newblock {\em Phys. Rev. Lett.}, 88:071805, 2002.

\bibitem{Hudson:2011zz}
J.J. Hudson, D.M. Kara, I.J. Smallman, B.E. Sauer, M.R. Tarbutt, et~al.
\newblock {Improved measurement of the shape of the electron}.
\newblock {\em Nature}, 473:493--496, 2011.

\bibitem{Liu:2007zf}
C.-P. Liu, M.J. Ramsey-Musolf, W.C. Haxton, R.G.E. Timmermans, and A.E.L.
  Dieperink.
\newblock {Atomic Electric Dipole Moments: The Schiff Theorem and Its
  Corrections}.
\newblock {\em Phys.Rev.}, C76:035503, 2007.

\bibitem{PhysRevA.77.014101}
R.~A. Sen'kov, N.~Auerbach, V.~V. Flambaum, and V.~G. Zelevinsky.
\newblock Reexamination of the schiff theorem.
\newblock {\em Phys. Rev. A}, 77:014101, 2008.

\bibitem{Flambaum:2001gq}
V.V. Flambaum and J.S.M. Ginges.
\newblock {The Nuclear Schiff moment and time invariance violation in atoms}.
\newblock {\em Phys.Rev.}, A65:032113, 2002.

\bibitem{Flambaum:2012vz}
V.V. Flambaum and A.~Kozlov.
\newblock {Screening and finite size corrections to the octupole and Schiff
  moments}.
\newblock {\em Phys.Rev.}, C85:068502, 2012.

\bibitem{Griffith:2009zz}
W.~C. Griffith et~al.
\newblock {Improved Limit on the Permanent Electric Dipole Moment of Hg-199}.
\newblock {\em Phys. Rev. Lett.}, 102:101601, 2009.

\bibitem{Latha:2009nq}
K.~V.~P. Latha, D.~Angom, B.~P. Das, and D.~Mukherjee.
\newblock {Probing CP violation with the electric dipole moment of atomic
  mercury}.
\newblock {\em Phys. Rev. Lett.}, 103:083001, 2009.

\bibitem{Dilipprivcomm}
Dilip Angom, priv. comm.

\bibitem{Dzuba:2002kg}
V.~A. Dzuba, V.~V. Flambaum, J.~S.~M. Ginges, and M.~G. Kozlov.
\newblock {Electric dipole moments of Hg, Xe, Rn, Ra, Pu, and TlF induced by
  the nuclear Schiff moment and limits on time- reversal violating
  interactions}.
\newblock {\em Phys. Rev.}, A66:012111, 2002.

\bibitem{Flambaum:1985ty}
V.~V. Flambaum, I.~B. Khriplovich, and O.~P. Sushkov.
\newblock Limit on the constant of t nonconserving nucleon nucleon interaction.
\newblock {\em Phys. Lett.}, B162:213--216, 1985.

\bibitem{Barton:1961eg}
G.~Barton.
\newblock {Notes on the static parity nonconserving internucleon potential}.
\newblock {\em Nuovo Cim.}, 19:512--527, 1961.

\bibitem{Ban:2010ea}
Shufang Ban, Jacek Dobaczewski, Jonathan Engel, and A.~Shukla.
\newblock {Fully self-consistent calculations of nuclear Schiff moments}.
\newblock {\em Phys. Rev.}, C82:015501, 2010.

\bibitem{Pospelov:2001ys}
Maxim Pospelov.
\newblock {Best values for the CP odd meson nucleon couplings from
  supersymmetry}.
\newblock {\em Phys.Lett.}, B530:123--128, 2002.

\bibitem{Bugg:2004cm}
D.~V. Bugg.
\newblock {The pion nucleon coupling constant}.
\newblock {\em Eur. Phys. J.}, C33:505--509, 2004.

\bibitem{Dmitriev:2004fk}
V.~F. Dmitriev, R.~A. Sen'kov, and N.~Auerbach.
\newblock {Effects of core polarization on the nuclear Schiff moment}.
\newblock {\em Phys. Rev.}, C71:035501, 2005.

\bibitem{deJesus:2005nb}
J.~H. de~Jesus and J.~Engel.
\newblock {Time-Reversal-Violating Schiff Moment of 199Hg}.
\newblock {\em Phys. Rev.}, C72:045503, 2005.

\bibitem{Shifman:1978zn}
Mikhail~A. Shifman, A.I. Vainshtein, and Valentin~I. Zakharov.
\newblock {Remarks on Higgs Boson Interactions with Nucleons}.
\newblock {\em Phys.Lett.}, B78:443, 1978.

\bibitem{Anselm:1985cf}
A.~A. Anselm, V.~E. Bunakov, Vladimir~P. Gudkov, and N.~G. Uraltsev.
\newblock On the neutron electric dipole moment in the weinberg cp violation
  model.
\newblock {\em Phys. Lett.}, B152:116--120, 1985.
\newblock [JETP Lett.40:1102-1105,1984].

\bibitem{Beringer:1900zz}
J.~Beringer et~al.
\newblock {Review of Particle Physics (RPP)}.
\newblock {\em Phys.Rev.}, D86:010001, 2012.

\bibitem{Alarcon:2011zs}
J.M. Alarcon, J.~Martin~Camalich, and J.A. Oller.
\newblock {The chiral representation of the $\pi N$ scattering amplitude and
  the pion-nucleon sigma term}.
\newblock {\em Phys.Rev.}, D85:051503, 2012.

\bibitem{Young:2009zb}
R.~D. Young and A.~W. Thomas.
\newblock {Octet baryon masses and sigma terms from an SU(3) chiral
  extrapolation}.
\newblock {\em Phys. Rev.}, D81:014503, 2010.

\bibitem{Toussaint:2009pz}
D.~Toussaint and W.~Freeman.
\newblock {The strange quark condensate in the nucleon in 2+1 flavor QCD}.
\newblock {\em Phys. Rev. Lett.}, 103:122002, 2009.

\bibitem{Takeda:2010cw}
K.~Takeda et~al.
\newblock {Nucleon strange quark content from two-flavor lattice QCD with exact
  chiral symmetry}.
\newblock {\em Phys. Rev.}, D83:114506, 2011.

\bibitem{Durr:2011mp}
S.~Durr, Z.~Fodor, T.~Hemmert, C.~Hoelbling, J.~Frison, et~al.
\newblock {Sigma term and strangeness content of octet baryons}.
\newblock {\em Phys.Rev.}, D85:014509, 2012.
\newblock Corrected error in the normalization of dimensionless sigma terms.

\bibitem{Ohki:2009mt}
H.~Ohki et~al.
\newblock {Nucleon sigma term and strange quark content in 2+1-flavor QCD with
  dynamical overlap fermions}.
\newblock {\em PoS}, LAT2009:124, 2009.

\bibitem{MartinCamalich:2010fp}
J.~Martin~Camalich, L.~S. Geng, and M.~J. Vicente~Vacas.
\newblock {The lowest-lying baryon masses in covariant SU(3)-flavor chiral
  perturbation theory}.
\newblock {\em Phys. Rev.}, D82:074504, 2010.

\bibitem{Bali:2011ks}
Gunnar~S. Bali et~al.
\newblock {The strange and light quark contributions to the nucleon mass from
  Lattice QCD}.
\newblock {\em Phys.Rev.}, D85:054502, 2012.

\bibitem{Dinter:2012tt}
Simon Dinter et~al.
\newblock {Sigma terms and strangeness content of the nucleon with $N_f=2+1+1$
  twisted mass fermions}.
\newblock {\em JHEP}, 1208:037, 2012.

\bibitem{Semke:2012gs}
A.~Semke and M.F.M. Lutz.
\newblock {Strangeness in the baryon ground states}.
\newblock {\em Phys.Lett.}, B717:242--247, 2012.

\bibitem{Borasoy:1996bx}
B.~Borasoy and Ulf-G. Meissner.
\newblock {Chiral expansion of baryon masses and sigma-terms}.
\newblock {\em Annals Phys.}, 254:192--232, 1997.

\bibitem{Dmitriev:2003sc}
V.F. Dmitriev and R.A. Sen'kov.
\newblock {Schiff moment of the mercury nucleus and the proton dipole moment}.
\newblock {\em Phys.Rev.Lett.}, 91:212303, 2003.

\bibitem{Baker:2006ts}
C.~A. Baker et~al.
\newblock {An improved experimental limit on the electric dipole moment of the
  neutron}.
\newblock {\em Phys. Rev. Lett.}, 97:131801, 2006.

\bibitem{1402-4896-36-3-011}
Ann-Marie M\aa{}rtensson-Pendrill and Per \"Oster.
\newblock Calculations of atomic electric dipole moments.
\newblock {\em Physica Scripta}, 36(3):444, 1987.

\bibitem{Flambaum:1985gx}
V.~V. Flambaum and I.~B. Khriplovich.
\newblock New limits on the electron dipole moment and t nonconserving electron
  - nucleon interaction.
\newblock {\em Sov. Phys. JETP}, 62:872--875, 1985.
\newblock [Zh.Eksp.Teor.Fiz.89:1505-1511,1985].

\bibitem{Kozlov:1988qn}
M.~G. Kozlov.
\newblock New limit on the scalar p, t odd electron nucleus interaction.
\newblock {\em Phys. Lett.}, A130:426--428, 1988.

\bibitem{Degrassi:2005zd}
Giuseppe Degrassi, Enrico Franco, Schedar Marchetti, and Luca Silvestrini.
\newblock {QCD corrections to the electric dipole moment of the neutron in the
  MSSM}.
\newblock {\em JHEP}, 11:044, 2005.

\bibitem{An:2009zh}
Haipeng An, Xiangdong Ji, and Fanrong Xu.
\newblock {P-odd and CP-odd Four-Quark Contributions to Neutron EDM}.
\newblock {\em JHEP}, 1002:043, 2010.

\bibitem{Hisano:2012cc}
Junji Hisano, Koji Tsumura, and Masaki~J.S. Yang.
\newblock {QCD Corrections to Neutron Electric Dipole Moment from Dimension-six
  Four-Quark Operators}.
\newblock {\em Phys.Lett.}, B713:473--480, 2012.

\bibitem{Dai:1989yh}
Jin Dai and Hans Dykstra.
\newblock {QCD corrections to CP violation in Higgs exchange}.
\newblock {\em Phys. Lett.}, B237:256, 1990.

\bibitem{Braaten:1990gq}
Eric Braaten, Chong-Sheng Li, and Tzu-Chiang Yuan.
\newblock The evolution of weinberg's gluonic cp violation operator.
\newblock {\em Phys. Rev. Lett.}, 64:1709, 1990.

\bibitem{Shifman:1976de}
Mikhail~A. Shifman, A.~I. Vainshtein, and Valentin~I. Zakharov.
\newblock {On the Weak Radiative Decays (Effects of Strong Interactions at
  Short Distances)}.
\newblock {\em Phys. Rev.}, D18:2583--2599, 1978.
\newblock [Erratum-ibid.D19:2815,1979].

\bibitem{Boyd:1990bx}
Glenn Boyd, Arun~K. Gupta, Sandip~P. Trivedi, and Mark~B. Wise.
\newblock Effective hamiltonian for the electric dipole moment of the neutron.
\newblock {\em Phys. Lett.}, B241:584, 1990.

\bibitem{2013PhRvA..87a2102S}
M.~D. {Swallows}, T.~H. {Loftus}, W.~C. {Griffith}, B.~R. {Heckel}, E.~N.
  {Fortson}, and M.~V. {Romalis}.
\newblock {Techniques used to search for a permanent electric dipole moment of
  the $^{199}$Hg atom and the implications for CP violation}.
\newblock {\em Phys. Rev. A}, 87(1):012102, 2013.

\bibitem{vanderGrinten:2009zz}
M.G.D. van~der Grinten et~al.
\newblock {CryoEDM: A cryogenic experiment to measure the neutron electric
  dipole moment}.
\newblock {\em Nucl.Instrum.Meth.}, A611:129--132, 2009.
\newblock See also {\tt https://www.neutronedm.org/index.html}.

\bibitem{SNSnEDM}
SNS nEDM Collaboration.
\newblock {\tt http://www.phy.ornl.gov/nedm/}.

\bibitem{Altarev:2009zz}
I.~Altarev, G.~Ban, G.~Bison, K.~Bodek, M.~Burghoff, et~al.
\newblock {Towards a new measurement of the neutron electric dipole moment}.
\newblock {\em Nucl.Instrum.Meth.}, A611:133--136, 2009.

\bibitem{Altarev:2012uy}
I.~Altarev, D.H. Beck, S.~Chesnevskaya, T.~Chupp, W.~Feldmeier, et~al.
\newblock {A next generation measurement of the electric dipole moment of the
  neutron at the FRM II}.
\newblock {\em Nuovo Cim.}, C035N04:122--127, 2012.
\newblock See also {\tt http://nedm.ph.tum.de/}.

\bibitem{Masuda20121347}
Yasuhiro Masuda et~al.
\newblock Neutron electric dipole moment measurement with a buffer gas
  comagnetometer.
\newblock {\em Physics Letters A}, 376(16):1347 -- 1351, 2012.

\bibitem{Serebrov2009263}
A.P. Serebrov et~al.
\newblock {Ultracold-neutron infrastructure for the PNPI/ILL neutron EDM
  experiment}.
\newblock {\em Nucl.Instrum.Meth.}, A611:263 -- 266, 2009.

\bibitem{Amini:2007ku}
Jason~M. Amini, Charles~T. Munger, Jr., and Harvey Gould.
\newblock {Electron electric dipole moment experiment using electric- field
  quantized slow cesium atoms}.
\newblock {\em Int. J. Mod. Phys.}, D16:2337--2342, 2008.
\newblock See also \texttt{http://homepage.mac.com/gould137/index.html}.

\bibitem{2004APS..DMP.P1056K}
M.~{Kittle}, T.~{Burton}, L.~{Feeney}, and D.~J. {Heinzen}.
\newblock {New experiment to measure the electron electric dipole moment}.
\newblock {\em APS Division of Atomic, Molecular and Optical Physics Meeting
  Abstracts}, pages 1056P--+, 2004.

\bibitem{Weissetal}
D.S. Weiss, F.~Fang, and J.~Chen.
\newblock {Measuring the electric dipole moment of Cs and Rb in an optical
  lattice}.
\newblock {\em Bull.Am.Phys.Soc.}, APR03:J1.008, 2003.
\newblock For the present status, see the talk by D. Weiss at the ``EDM
  searches at Storage Rings'' workshop at ECT*, {\tt http://www.ectstar.eu}.

\bibitem{Sakemi:2011zz}
Y.~Sakemi, K.~Harada, T.~Hayamizu, M.~Itoh, H.~Kawamura, et~al.
\newblock {Search for a permanent EDM using laser cooled radioactive atom}.
\newblock {\em J.Phys.Conf.Ser.}, 302:012051, 2011.

\bibitem{PhysRevX.2.041009}
B.~J. Wundt, C.~T. Munger, and U.~D. Jentschura.
\newblock Quantum dynamics in atomic-fountain experiments for measuring the
  electric dipole moment of the electron with improved sensitivity.
\newblock {\em Phys. Rev. X}, 2:041009, 2012.
\newblock See also {\tt http://eedm.info/index.html}.

\bibitem{1367-2630-15-5-053034}
M~R Tarbutt, B~E Sauer, J~J Hudson, and E~A Hinds.
\newblock Design for a fountain of ybf molecules to measure the electron's
  electric dipole moment.
\newblock {\em New Journal of Physics}, 15(5):053034, 2013.

\bibitem{Vutha:2009ux}
Amar~C. Vutha, Wesley~C. Campbell, Yulia~V. Gurevich, Nicholas~R. Hutzler,
  Maxwell Parsons, et~al.
\newblock {Search for the electric dipole moment of the electron with thorium
  monoxide}.
\newblock {\em J.Phys.}, B43:074007, 2010.

\bibitem{HgExp}
Information taken from http://nedm.web.psi.ch/EDM-world-wide/; see also
  http://oldwww.phys.washington.edu/users/fortson/.

\bibitem{PhysRevLett.86.22}
M.~A. Rosenberry and T.~E. Chupp.
\newblock Atomic electric dipole moment measurement using spin exchange pumped
  masers of ${}^{129}\mathrm{Xe}$ and ${}^{3}\mathrm{He}$.
\newblock {\em Phys. Rev. Lett.}, 86:22--25, 2001.

\bibitem{Gemmel:2009pu}
C.~Gemmel, W.~Heil, K.~Lenz, Ch. Ludwig, K.~Thulley, et~al.
\newblock {Ultra-sensitive magnetometry based on free precession of nuclear
  spins}.
\newblock {\em Eur.Phys.J.}, D57:303--320, 2010.

\bibitem{XenonExp}
See e.g. http://cns.pnpi.spb.ru/ucn/articles/Taubenheim.pdf.

\bibitem{Gaffney:2013xx}
L.P. Gaffney et~al.
\newblock Studies of pear-shaped nuclei using accelerated radioactive beams.
\newblock {\em Nature}, 497:199, 2013.

\bibitem{Serebrov:2013tba}
A.P. Serebrov, E.A. Kolomenskiy, A.N. Pirozhkov, I.A. Krasnoshekova, A.V.
  Vasiliev, et~al.
\newblock {New measurements of neutron electric dipole moment}.
\newblock 2013.
\newblock arXiv:1310.5588 [nucl-ex].

\bibitem{2013PhRvA..87e2130E}
S.~{Eckel}, P.~{Hamilton}, E.~{Kirilov}, H.~W. {Smith}, and D.~{DeMille}.
\newblock {Search for the electron electric dipole moment using
  {$\Omega$}-doublet levels in PbO}.
\newblock {\em Phys. Rev. A}, 87(5):052130, 2013.

\bibitem{Lamoreaux:2001hb}
S.K. Lamoreaux.
\newblock {Solid state systems for electron electric dipole moment and other
  fundamental measurements}.
\newblock {\em Phys.Rev.}, A66:022109, 2002.

\bibitem{0038-5670-11-3-A11}
F~L Shapiro.
\newblock Electric dipole moments of elementary particles.
\newblock {\em Soviet Physics Uspekhi}, 11(3):345, 1968.

\bibitem{Eckel:2012aw}
S.~Eckel, A.O. Sushkov, and S.K. Lamoreaux.
\newblock {A limit on the electron electric dipole moment using paramagnetic
  ferroelectric Eu$_{0.5}$Ba$_{0.5}$TiO$_3$}.
\newblock {\em Phys.Rev.Lett.}, 109:193003, 2012.

\bibitem{Cho:1989hd}
D.~Cho, K.~Sangster, and E.A. Hinds.
\newblock Tenfold improvement of limits on t violation in thallium fluoride.
\newblock {\em Phys.Rev.Lett.}, 63:2559--2562, 1989.

\bibitem{Garwin:1959xx}
R.L. Garwin and L.~Lederman.
\newblock The electric dipole moment of elementary particles.
\newblock {\em Nuovo Cim.}, 11:776, 1959.

\bibitem{Semertzidis:1998sp}
Y.~K. Semertzidis.
\newblock {A new experiment for an electric dipole moment of muon at the
  10**(-24)-e-cm level}.
\newblock In E.~Zavattini, D.~Bakalov, and C.~Rizzo, editors, {\em Frontier
  tests of QED and physics of the vacuum. Proceedings, Workshop, Sandansky,
  Bulgaria, June 9-15, 1998}, 1998.

\bibitem{Khriplovich:1998zq}
I.~B. Khriplovich.
\newblock {Feasibility of search for nuclear electric dipole moments at ion
  storage rings}.
\newblock {\em Phys. Lett.}, B444:98--102, 1998.

\bibitem{Farley:2003wt}
F.~J.~M. Farley et~al.
\newblock {A new method of measuring electric dipole moments in storage rings}.
\newblock {\em Phys. Rev. Lett.}, 93:052001, 2004.

\bibitem{Bennett:2008dy}
G.W. Bennett et~al.
\newblock {An Improved Limit on the Muon Electric Dipole Moment}.
\newblock {\em Phys.Rev.}, D80:052008, 2009.

\bibitem{Semertzidis:2011qv}
Yannis~K. Semertzidis.
\newblock {A Storage Ring proton Electric Dipole Moment experiment: most
  sensitive experiment to CP-violation beyond the Standard Model}.
\newblock 2011.
\newblock arXiv:1110.3378 [hep-ph].

\bibitem{Kawall:2011zz}
D.~Kawall.
\newblock {Searching for the electron EDM in a storage ring}.
\newblock {\em J.Phys.Conf.Ser.}, 295:012031, 2011.
\newblock See also talk at the ${\rm ECT}^*$ Storage Ring EDM Workshop, October
  2012.

\bibitem{Lebedev:2002ne}
Oleg Lebedev and Maxim Pospelov.
\newblock {Electric dipole moments in the limit of heavy superpartners}.
\newblock {\em Phys.Rev.Lett.}, 89:101801, 2002.

\bibitem{Shabalin:1978rs}
E.P. Shabalin.
\newblock {Electric Dipole Moment of Quark in a Gauge Theory with Left-Handed
  Currents}.
\newblock {\em Sov.J.Nucl.Phys.}, 28:75, 1978.

\bibitem{Barr:1990vd}
Stephen~M. Barr and A.~Zee.
\newblock Electric dipole moment of the electron and of the neutron.
\newblock {\em Phys. Rev. Lett.}, 65:21--24, 1990.
\newblock Erratum-ibid.65:2920,1990.

\bibitem{Gunion:1990iv}
J.~F. Gunion and D.~Wyler.
\newblock {Inducing a large neutron electric dipole moment via a quark
  chromoelectric dipole moment}.
\newblock {\em Phys. Lett.}, B248:170--176, 1990.

\bibitem{Chang:1990twa}
D.~Chang, Wai-Yee Keung, and T.~C. Yuan.
\newblock {Chromoelectric dipole moment of light quarks through two loop
  mechanism}.
\newblock {\em Phys. Lett.}, B251:608--612, 1990.

\bibitem{BowserChao:1997bb}
David Bowser-Chao, Darwin Chang, and Wai-Yee Keung.
\newblock {Electron electric dipole moment from CP violation in the charged
  Higgs sector}.
\newblock {\em Phys. Rev. Lett.}, 79:1988--1991, 1997.

\bibitem{Chang:1999zw}
Darwin Chang, We-Fu Chang, and Wai-Yee Keung.
\newblock {Additional two-loop contributions to electric dipole moments in
  supersymmetric theories}.
\newblock {\em Phys. Lett.}, B478:239--246, 2000.

\bibitem{Pilaftsis:1999td}
Apostolos Pilaftsis.
\newblock {Higgs-boson two-loop contributions to electric dipole moments in the
  MSSM}.
\newblock {\em Phys. Lett.}, B471:174--181, 1999.

\bibitem{Pospelov:1991zt}
M.E. Pospelov and I.B. Khriplovich.
\newblock {Electric dipole moment of the W boson and the electron in the
  Kobayashi-Maskawa model}.
\newblock {\em Sov.J.Nucl.Phys.}, 53:638--640, 1991.
\newblock Yad.Fiz. 53 (1991) 1030-1033.

\bibitem{Pich:2013vta}
Antonio Pich.
\newblock {The Physics of the Higgs-like Boson}.
\newblock {\em EPJ Web Conf.}, 60:02006, 2013.

\bibitem{Braeuninger:2010td}
Carolin~B. Braeuninger, Alejandro Ibarra, and Cristoforo Simonetto.
\newblock {Radiatively induced flavour violation in the general two-Higgs
  doublet model with Yukawa alignment}.
\newblock {\em Phys.Lett.}, B692:189--195, 2010.

\bibitem{Ferreira:2010xe}
P.M. Ferreira, L.~Lavoura, and Joao~P. Silva.
\newblock {Renormalization-group constraints on Yukawa alignment in
  multi-Higgs-doublet models}.
\newblock {\em Phys.Lett.}, B688:341--344, 2010.

\bibitem{Dicus:1989va}
Duane~A. Dicus.
\newblock Neutron electric dipole moment from charged higgs exchange.
\newblock {\em Phys. Rev.}, D41:999, 1990.

\bibitem{Grinstein:1987vj}
Benjamin Grinstein, Roxanne~P. Springer, and Mark~B. Wise.
\newblock {Effective Hamiltonian for Weak Radiative B Meson Decay}.
\newblock {\em Phys. Lett.}, B202:138, 1988.

\bibitem{DeRujula:1990wy}
A.~De~Rujula, M.~B. Gavela, O.~Pene, and F.~J. Vegas.
\newblock {Even larger contributions to the neutron electric dipole moment}.
\newblock {\em Phys. Lett.}, B245:640--648, 1990.

\bibitem{Gunion:1990ce}
J.F. Gunion and R.~Vega.
\newblock {The Electron electric dipole moment for a CP violating neutral Higgs
  sector}.
\newblock {\em Phys.Lett.}, B251:157--162, 1990.

\bibitem{Leigh:1990kf}
R.G. Leigh, S.~Paban, and R.M. Xu.
\newblock {Electric dipole moment of electron}.
\newblock {\em Nucl.Phys.}, B352:45--58, 1991.

\bibitem{Chang:1990sf}
D.~Chang, Wai-Yee Keung, and T.C. Yuan.
\newblock {Two loop bosonic contribution to the electron electric dipole
  moment}.
\newblock {\em Phys.Rev.}, D43:14--16, 1991.

\bibitem{Kao:1992jv}
Chung Kao and Rui-Ming Xu.
\newblock {Charged Higgs loop contribution to the electric dipole moment of
  electron}.
\newblock {\em Phys.Lett.}, B296:435--439, 1992.

\end{thebibliography}
\end{document}